\documentclass[10pt,conference]{IEEEtran}

\usepackage{cite}
\usepackage{amsmath,amssymb,amsfonts}
\usepackage{graphicx}
\usepackage{textcomp}
\usepackage[hyphens]{url}

\usepackage{fancyhdr}
\usepackage[hidelinks]{hyperref}
\usepackage{braket}
\usepackage{subcaption}
\usepackage{booktabs}
\usepackage{makecell}
\usepackage{float}
\usepackage{enumitem}
\usepackage{multirow}
\usepackage[x11names]{xcolor}
\makeatletter
\newcommand{\linebreakand}{
  \end{@IEEEauthorhalign}
  \hfill\mbox{}\par
  \mbox{}\hfill\begin{@IEEEauthorhalign}
}
\makeatother
\newcommand{\Rz}{\texorpdfstring{$R_z$}{Rz}}
\usepackage{tikz}
\usetikzlibrary{svg.path}
\usepackage{scalerel}

\begin{document}
\definecolor{orcidlogocol}{HTML}{A6CE39}
\tikzset{
  orcidlogo/.pic={
    \fill[orcidlogocol]
    svg{M256,128c0,70.7-57.3,128-128,128C57.3,256,0,198.7,0,128C0,57.3,57.3,0,128,0C198.7,0,256,57.3,256,128z};
    \fill[white] svg{M86.3,186.2H70.9V79.1h15.4v48.4V186.2z}
    svg{M108.9,79.1h41.6c39.6,0,57,28.3,57,53.6c0,27.5-21.5,53.6-56.8,53.6h-41.8V79.1z
    M124.3,172.4h24.5c34.9,0,42.9-26.5,42.9-39.7c0-21.5-13.7-39.7-43.7-39.7h-23.7V172.4z}
    svg{M88.7,56.8c0,5.5-4.5,10.1-10.1,10.1c-5.6,0-10.1-4.6-10.1-10.1c0-5.6,4.5-10.1,10.1-10.1C84.2,46.7,88.7,51.3,88.7,56.8z};
  }
}

\newcommand\orcidicon[1]{\textsuperscript{\href{https://orcid.org/#1}{\mbox{\scalerel*{
          \begin{tikzpicture}[yscale=-1,transform shape]
            \pic{orcidlogo};
          \end{tikzpicture}
}{|}}}}}

\title{Price and Payoff: Non-Determinism in Fault Tolerant Quantum Computation}
\author{
\IEEEauthorblockN{
Aditi Awasthi\IEEEauthorrefmark{3}\IEEEauthorrefmark{1}\orcidicon{0009-0006-7768-8503},
Sayam Sethi\IEEEauthorrefmark{3}\orcidicon{0009-0005-3056-5285},
Sahil Khan\IEEEauthorrefmark{2}\orcidicon{0009-0000-4160-8010},
Gokul Subramanian Ravi\IEEEauthorrefmark{4}\orcidicon{0000-0002-2334-2682},
Jonathan Mark Baker\IEEEauthorrefmark{3}\orcidicon{0000-0002-0775-8274}
}
\IEEEauthorblockA{
\IEEEauthorrefmark{3}Electrical and Computer Engineering, The University of Texas at Austin\\
\IEEEauthorrefmark{2}Electrical and Computer Engineering, Duke University\\
\IEEEauthorrefmark{4}Computer Science and Engineering, University of Michigan\\
\IEEEauthorrefmark{1}\href{mailto:aditiawasthi@utexas.edu}{aditiawasthi@utexas.edu}
}
}
\maketitle

\begin{abstract}
A promising approach to achieving scalable fault-tolerant quantum computation (FTQC) is the use of quantum error correction (QEC) codes augmented with magic states i.e. resource states produced via distillation~\cite{bravyi2012magic, Litinski_2019_distillation}, cultivation~\cite{Gidney_2024_cultivation}, or \Rz{} synthesis~\cite{akahoshi2024partially,yoshioka_transversal_2025} and teleported into the circuit as needed. Because magic-state production dominates the space-time volume of fault-tolerant programs, system architects must decide how many production units to allocate. Current approaches~\cite{qre1,qre2} rely on deterministic analysis that either provisions for worst-case peak demand (wasting valuable qubit resources on factories that are never simultaneously utilized) or assumes average demand, which increases execution time.

In this work, we build a simulation framework that couples circuit scheduling with different stochastic magic state production models, and use it to quantify the impact of non-determinism on circuit execution. We show that non-determinism has a dual effect that deterministic models cannot capture: it \emph{inflates} total execution time (the price), while \emph{deflating} peak per-cycle resource demand (the payoff). For distillation-based architectures, this demand smoothing shifts the space-time-optimal provisioning point: fewer factories are needed to minimize space-time volume than deterministic analysis predicts. Across benchmarks, stochastic-aware provisioning reduces space-time volume by up to 27\% compared to the deterministic optimum for distillation, while requiring up to 30\% fewer factories. We characterize these effects across each preparation mechanism, map the resulting design-space tradeoffs, and demonstrate that static resource estimation systematically mis-characterizes the cost of fault-tolerant execution. Our results establish that stochastic-aware analysis is necessary for right-sizing the factory allocations and should replace deterministic heuristics as the standard methodology for FTQC resource planning.
\end{abstract}

\begin{IEEEkeywords}
magic state distillation factories, magic state cultivation, \Rz{} synthesis, space-time volume, injection errors, factory aborts
\end{IEEEkeywords}

\section{Introduction}
Fault-tolerant quantum computing (FTQC) relies on quantum error correction (QEC) to enable reliable execution of quantum algorithms. QEC requires an enormous overhead: thousands to millions of physical qubits encode each logical qubit~\cite{gidney2021factor, terhal2015quantum, Litinski_2019_distillation}, making resource-efficient system design essential. Clifford gates (H, S, CNOT) can be implemented directly through code deformations~\cite{fowler_2012}, while T and \Rz{} gates require auxiliary ``magic states" produced by specialized distillation factories \cite{bravyi2012magic} or alternative methods such as cultivation~\cite{Gidney_2024_cultivation} or direct \Rz{} synthesis. These gates are crucial: circuits containing only Clifford gates can be simulated classically in polynomial time, making them an indirect measure of quantum computational advantage.  The magic resource states needed to implement these gates are the dominant contributor to both execution time and physical qubit requirements~\cite{Litinski_2019_lattice_surgery,holmes2019resource}, making the problem of determining how many production units to allocate, and of what type, a crucial architectural decisions in FTQC system design.

\begin{figure*}
    \centering
    \includegraphics[width=\linewidth]{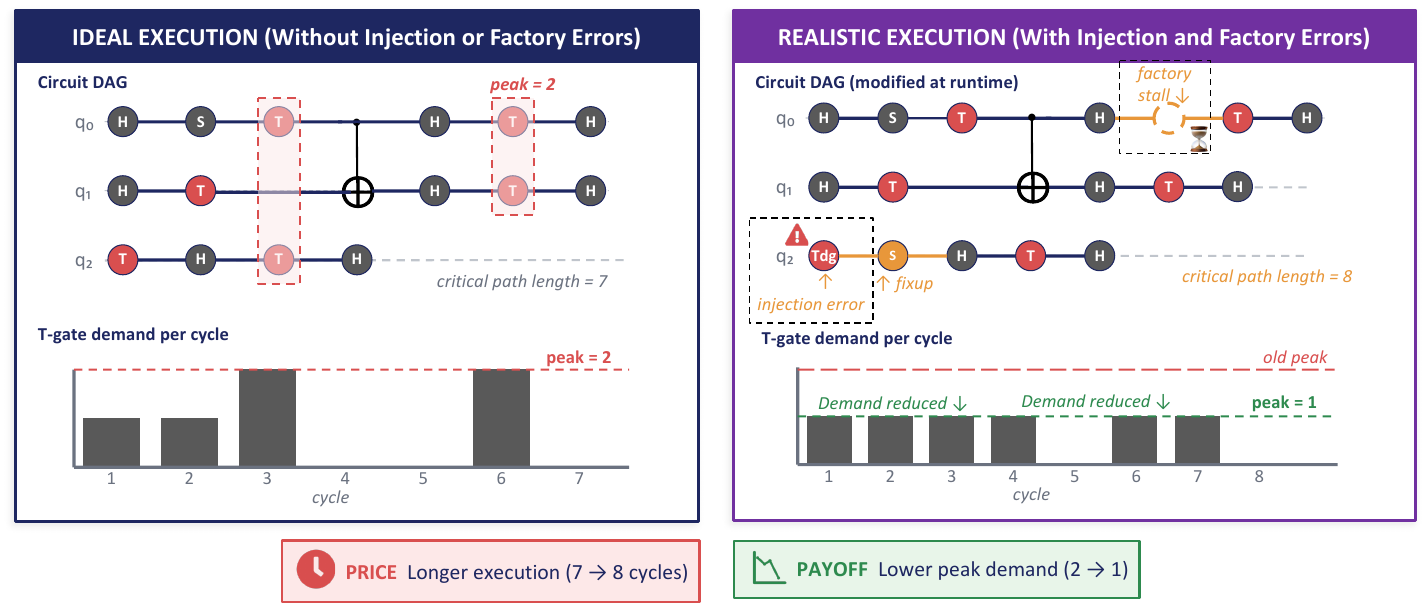}
    \caption{(Left) Circuit execution under deterministic assumptions and clean critical path. (Right) Same circuit subjected to non-deterministic factors: fix-up operations are inserted (highlighted in yellow). We observe that the T-gate demand gets stretched and flattened, hence the number of magic states needed per cycle is visibly lower}
    \label{fig:introfigure}
\end{figure*}

This allocation decision is difficult because magic-state demand is highly non-uniform, for example some circuit layers require resources on every qubit while others require none. Compounding this variability, magic-state production and consumption are both non-deterministic. Factory-level failures (distillation aborts, failed cultivation stages, variable \Rz{} preparation times) reduce effective throughput, while injection errors (which occur with $50\%$ probability on every T or \Rz{} gate) insert corrective fixup operations into the circuit at runtime, reshaping the execution schedule. These two stochastic effects interact with the circuit's dependency structure in ways that cannot be predicted from static analysis alone. Figure~\ref{fig:introfigure} illustrates this dual effect even on a small example: injection errors and factory stalls stretch the schedule from 7 to 8 cycles (the price), while flattening peak T-gate demand from 2 to 1 per cycle (the payoff). The interplay between these stochastic effects and the factory's spatial footprint, production latency, and output fidelity makes provisioning a multi-dimensional optimization problem that static analysis cannot adequately address.

Due to this complexity, existing approaches to factory provisioning have relied on deterministic analysis. Current theoretical work~\cite{qre1, qre2} assumes unlimited magic-state capacity or that states are prepared offline in advance, while practical resource estimators provision for worst-case demand at every time step. The former ignores throughput constraints entirely; the latter leads to systematic over-provisioning, and hence under-utilization of allocated quantum resources. Neither approach models the stochastic errors which slow execution during runtime and reshape the demand profile. This is perhaps unsurprising; deterministic worst-case provisioning provides a conservative estimate without needing to model the stochastic effects that complicate analysis. Our key insight is that these estimates are not merely cautious but systematically wasteful, because the same errors that justify caution also reshape demand in ways that reduce the required capacity.

Several prior works optimize the internal design of magic-state factory circuits~\cite{Ding_2018,holmes2019resource,chadwick2024averting,xu_distilling_2026}; we treat the factory as a black-box, making our approach easily extensible to novel factory designs proposed in literature~\cite{xu_distilling_2026,yoshioka_transversal_2025}. We address the complementary (and previously unexplored) system-level problem of how many factories to allocate given realistic, stochastic production and consumption. 
We operate under a cold-start model (i.e. no pre-filled buffers), allowing us to observe the full online interaction between supply variability and demand dynamics, including the transient starvation period at the start of execution.

Our contributions include:
\begin{itemize}
    \item A simulation framework coupling critical-path-priority list scheduling with stochastic production models for three mechanisms: 15-to-1 distillation, magic-state cultivation, and direct \Rz{} synthesis. Our approach can also easily be extended to novel factory designs.
    \item Quantification of the \emph{price} of non-determinism: factory errors and injection-error fix-ups inflate execution time by up to $2.5\times$ for cultivation and up to $1.4\times$ for distillation relative to deterministic predictions, depending on provisioning level and circuit structure.
    \item Discovery and quantification of the \emph{payoff}: stochastic execution reduces the realized peak per-cycle magic-state demand below the deterministic peak (e.g. from 15 to 13 T gates per cycle on knn\_n25 in case of distillation), enabling fewer factories without incurring a loss in throughput.
    \item Design-space exploration showing that the space-time-optimal factory count under stochastic execution differs from (and for distillation is strictly lower than) the deterministic optimum,
    yielding up to 27\% space-time savings across benchmarks, and demonstrating that static resource estimation systematically mis-allocates qubit budget.
\end{itemize}

\section{Background}
\subsection{Eastin-Knill Theorem and Universality}
Constructing a universal gate set for quantum error correcting codes is limited by the Eastin-Knill theorem~\cite{eastin2009restrictions}, which states that no quantum error correcting code can support a universal set of transversal logical gates. Consequently, surface code architectures typically implement the Clifford+T gate set, where the gates $\{$CNOT, H, S$\}$ can be performed fault-tolerantly (via lattice surgery). The T gate requires special handling through magic state injection (Section~\ref{sec:injection-models}). Other codes may require some other non-T gate to generate universality (for example using code switching \cite{kubica2015universal}). The Clifford+T gate set achieves universality through the ability to approximate arbitrary single-qubit rotations with sequences of discrete gates. Input circuits containing continuous rotations $R_z(\theta)$ are decomposed into sequences of H, S, and T gates~\cite{matsumoto_2008,giles_2019} using techniques like GridSynth~\cite{ross_2016_gridsynth}.

\begin{figure}
    \centering
    \includegraphics[width=1.05\linewidth]{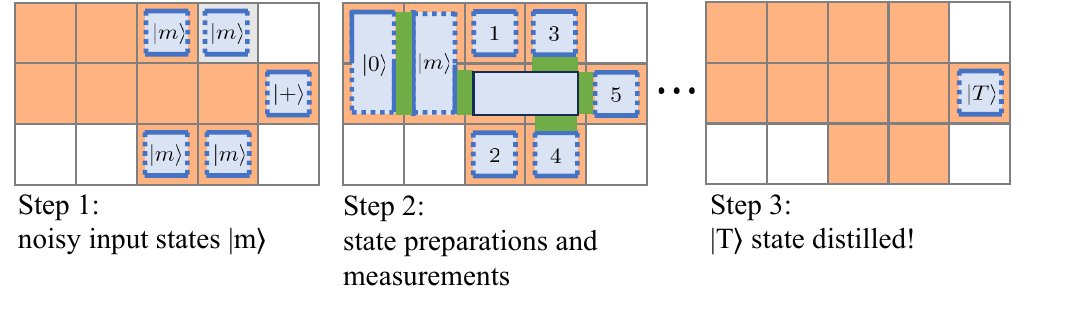}
    \caption{Resource state production using lattice surgery to perform magic state distillation in factories (here, a 15-to-1 factory. A factory transforms several noisy input states $\ket{m}$ (step 1) into a single higher quality $\ket{T}$ state (step 3) using a series of state preparations and measurements (step 2). The process takes 10s of cycles to complete \cite{Litinski_2019_distillation, chadwick2024averting}}
    \label{fig:surface-code-arch}
\end{figure}

\subsection{Significance of T count and Circuit Structure}
In practical settings, the T-gate count and circuit structure shape both algorithm design and architectural choices. T gates constitute almost 25\%--30\% of useful quantum applications \cite{TamingBW_2017}, and some application types claim even higher percentages (40\%--47\%) \cite{Isailovic_2008}. Consider Shor's algorithm for factoring an integer $N$ with bit-size $n$: the algorithm requires $2n + 2$ logical qubits and fewer than $448n^3 \log_2(n)$ T gates. For $n = 1024$, this translates to 2050 logical qubits and approximately $4.81 \times 10^{12}$ T gates~\cite{haner2017factoringusing2n2qubits}. At these scales, even modest improvements in factory provisioning efficiency translate to substantial qubit savings. Magic-state factories are consequently the rate-limiting component and dominant consumer of physical qubits in fault-tolerant architectures~\cite{holmes2019resource}. This makes factory provisioning, i.e. how many factories to build and of what type, the primary lever for trading off space against time in FTQC execution, and the central subject of this work.

\section{Stochastic Production and Consumption Models}\label{sec:injection-models}
Each magic-state preparation mechanism introduces non-determinism with distinct characteristics. In this section, we define the stochastic models used in our simulation framework, focusing on the failure modes and timing variability that drive the effects measured in further Sections~\ref{sec:price} and~\ref{sec:payoff}.

\subsection{Magic state distillation}
Magic-state distillation \cite{Litinski_2019_distillation} consumes multiple noisy magic states (and ancillary qubits) to probabilistically produce fewer, higher-fidelity outputs using fixed-depth logical Clifford circuits. This is implemented using dedicated factories that operate continuously in parallel with the main computation. A distillation factory is modeled as a $n$-to-$m$ factory where $n$ noisy magic states are distilled to produce $m$ high fidelity states. To meet the expected T-gate demand of the computation we need to select among different factory configurations, parametrized by $n$, $m$, program size, target fidelity, etc.
The spacetime cost of distillation is driven by large logical patches, multi-round stabilizer measurements, and lattice-surgery operations. In our model, factory failures manifest as throughput shortfalls. When a factory's production round fails, the computation must either 1) consume a state from another factory, 2) stall until the next successful round, or 3) draw from a buffer (if present). For example, the 15-to-1 protocol (shown in Figure~\ref{fig:surface-code-arch}) has a discard rate $p_{\mathrm{abort}} \approx 15p + 105p^{2}$~\cite{Bravyi_2005}, where $p$ is the physical error rate, roughly 0.15\% at $p=10^{-4}$. A failed round wastes a full production interval, delaying the next available state. Crucially, the factory's large spatial footprint implies that each provisioned but underutilized factory leads to a significant qubit overhead, whereas insufficient factory provisioning increases overall circuit execution time when magic-state demand exceeds supply.

\begin{figure}[!b]
    \centering
    \begin{subfigure}{\linewidth}
        \centering
        \includegraphics[width=0.9\linewidth]{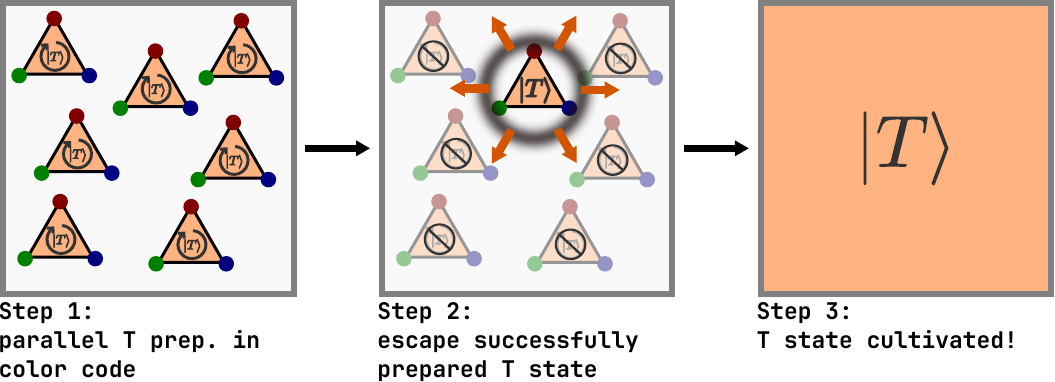}
        \caption{Abstracted depiction of cultivation where $\ket{T} = T\ket{+}$}\label{fig:cultivation_methodology}
        \vspace{6pt}
    \end{subfigure}
    \begin{subfigure}{\linewidth}
        \centering
        \includegraphics[width=0.9\linewidth]{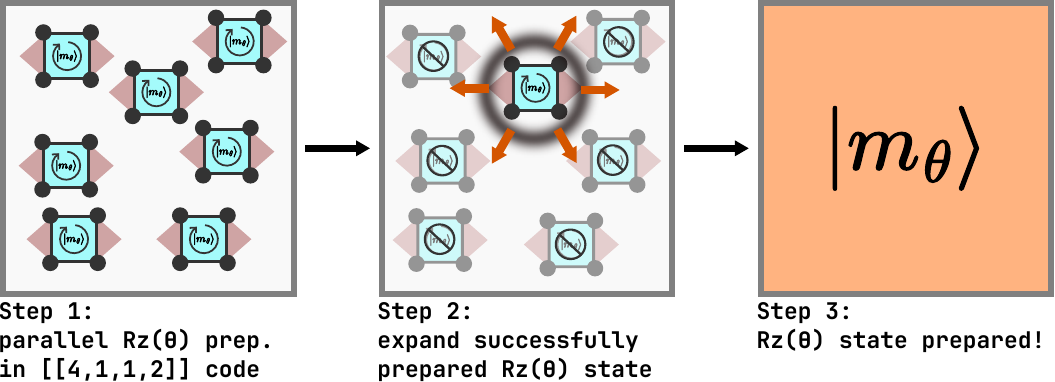}
        \caption{Abstracted depiction of \Rz{}-synthesis where $\ket{m_\theta} = Rz(\theta)\ket{+}$}\label{fig:rzsynth_methodology}
    \end{subfigure}
    \caption{Multiple state preparations are attempted in parallel on color codes (for cultivation) and $[[4,1,1,2]]$ code (for \Rz{} synthesis), and the successfully prepared state escapes/expands into the higher distance code. Each of the step can fail with non-zero probability which requires restarting from Step 1, irrespective of when the failure happened.}
\end{figure}

\begin{figure}
    \centering
    \begin{subfigure}{\linewidth}
    \centering
    \includegraphics[width=0.65\linewidth]{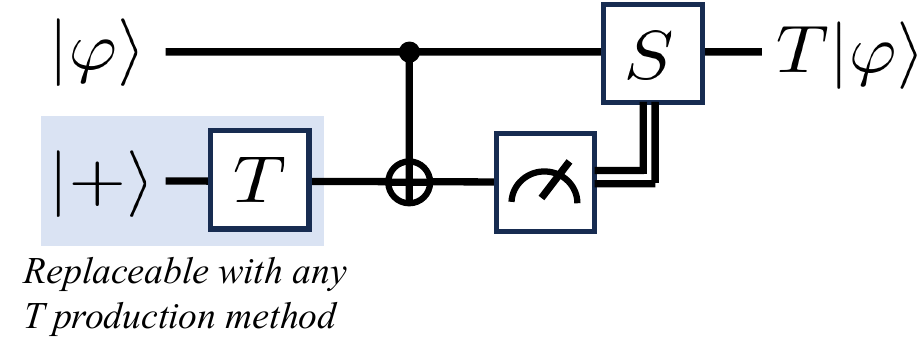}
    \caption{$\ket{T}$ injection onto $\ket{\varphi}$. The magic state is produced remotely, e.g. in a distillation factory or cultivation patch shown in blue}
    \end{subfigure}
    \hfill \vspace{2pt}
    \begin{subfigure}{\linewidth}
    \centering
    \includegraphics[width=0.8\linewidth]{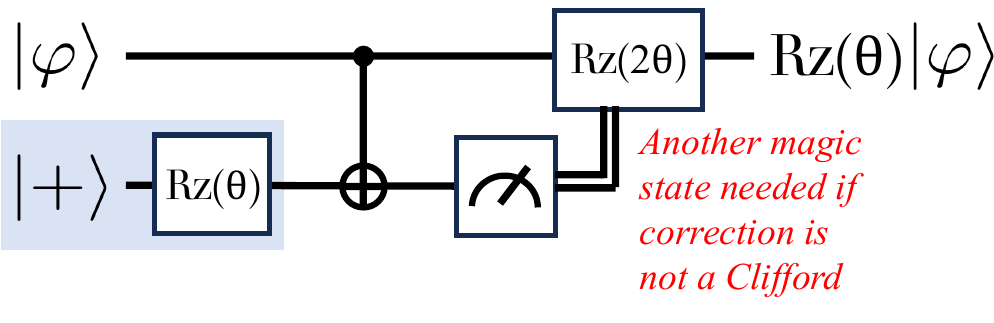}
    \caption{$\ket{R_z(\theta}$ injection onto $\ket{\varphi}$. The magic state is produced remotely in an \Rz{}-synthesis patch shown in blue}
    \end{subfigure}
    \caption{Magic state injection using teleportation, including conditional correction}
    \label{fig:gateinjection}
\end{figure}

\subsection{Magic state cultivation}
Magic-state cultivation \cite{Gidney_2024_cultivation} has been introduced as a more efficient approach to obtain magic states, which operates by incrementally increasing the fidelity of a magic state by repeatedly checking it and growing the distance of the code that hosts it. This is an intrinsically non-deterministic process: production proceeds through multiple check stages whose individual success probabilities depend on the code parameters $(d_1, d_2, r_1, r_2)$ and a failure at any stage restarts production from scratch. We illustrate the cultivation protocol in Figure~\ref{fig:cultivation_methodology}, where $d_1$ is the distance of the inner color code, $d_2$ is the distance of the outer escaped code, $r_1$ is the number of rounds spent in stage 1 after step 1 succeeds and $r_2$ is the number of rounds spent in stage 2 after step 2 succeeds. If either of the step fails, execution restarts from step 1. A usable output is produced only if the magic state successfully survives all stages i.e. the injection, cultivation, and escape stages, without triggering a discard condition during post-selection.
Cultivation introduces variability in production latency, in contrast to distillation's relatively predictable (albeit long) production intervals. The dominant cost shifts from large, persistent distillation factories to smaller, transient code patches whose size grows over time. Although qubit footprint for a single cultivation unit is much lower than that of a distillation factory, the expected spacetime cost per accepted state includes the cumulative cost of failed attempts. This high failure rate but fast retry characteristic produces a qualitatively different interaction than distillation, one where the the supply is unreliable on any given cycle but recovers quickly.

\subsection{Injection Errors}
The second source of non-determinism arises at the point of magic state consumption. When a magic state is consumed to implement a T gate, the injection correctly applies the desired T injection with 50\% probability. However, with 50\% probability, the injection fails and instead applies a $T^\dagger$ injection. In this case, correctness needs to be restored by dynamically applying an additional \emph{fixup} S gate (using the circuit identity $T^\dagger S = T$), as shown in Figure~\ref{fig:gateinjection}. Importantly, this recovery mechanism does not require the production of an additional magic state; instead, it is handled via dynamic modification of the circuit directly. While we describe this process here for T gates specifically, an analogous injection failure mechanism applies to $R_z(\theta)$ rotations; we detail this in the next subsection.
Crucially, each fixup operation needs to be completed before execution can proceed. This can extend the critical path length, delaying every downstream operation that transitively depends on the fixup. The scheduler must then recompute critical-paths, potentially reordering which gates are serviced first in subsequent cycles. This interaction between injection errors and scheduling is the primary driver of both the price and the payoff that we quantify in Sections~\ref{sec:price} and~\ref{sec:payoff}.

\subsection{\Rz{} synthesis}

The conventional approach to implementing $R_z(\theta)$ operations is to approximate them using long Clifford+T gate sequences, incurring substantial increases in depth and magic-state overhead. However, recent works \cite{akahoshi2024partially, Sethi_2025} have instead proposed realizing arbitrary $R_z(\theta)$ gates by preparing and injecting rotation resource states of the form $\ket{m_\theta} = R_z(\theta)\ket{+}$ using a repeat-until-success protocol. This approach directly implements the desired rotation without decomposing into T gates, significantly reducing the space-time cost of non-Clifford operations.
Each distinct rotation angle $\theta$ in the circuit requires a dedicated production unit that prepares the angle-specific resource state. The underlying production
protocol is a cultivation-like repeat-until-success
process (Figure~\ref{fig:rzsynth_methodology}); the preparation time per state depends on the code distance~$d$ and physical error rate~$p$ through a retry process over preparation-and-expansion rounds.

\begin{figure*}[!ht]
  \centering
  \begin{subfigure}{\textwidth}
    \centering
        \includegraphics[width=\linewidth]{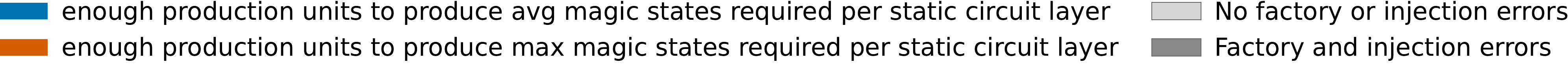}
    \end{subfigure}\\
    \vspace{2pt}
  \begin{subfigure}{\textwidth}
    \includegraphics[height=4cm,width=\columnwidth]{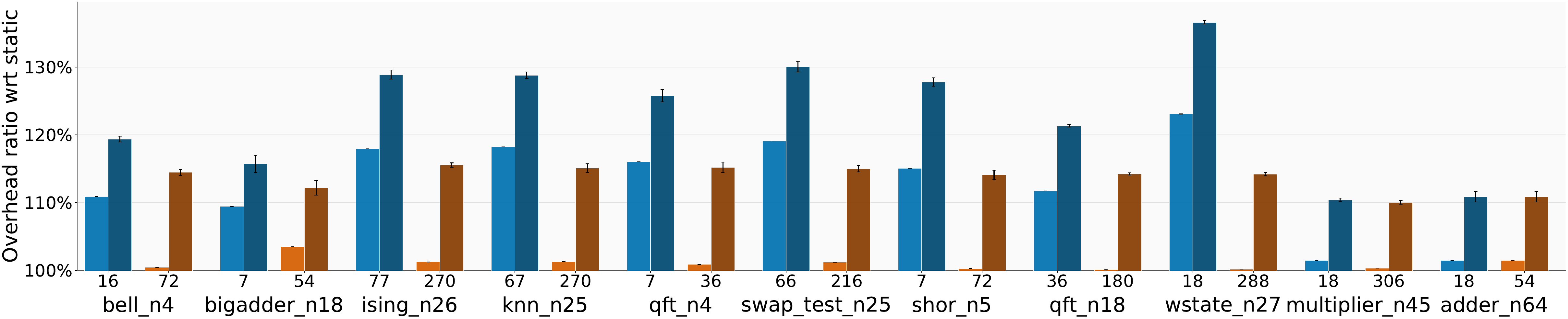}
    \caption{(15-to-1)$_{7,3,3}$ magic state distillation at 1e-4 physical error rate}
    \label{fig:overhead_cycles_distil}
  \end{subfigure}
  \hfill
  \begin{subfigure}{\textwidth}
    \includegraphics[height=3.7cm,width=\columnwidth]{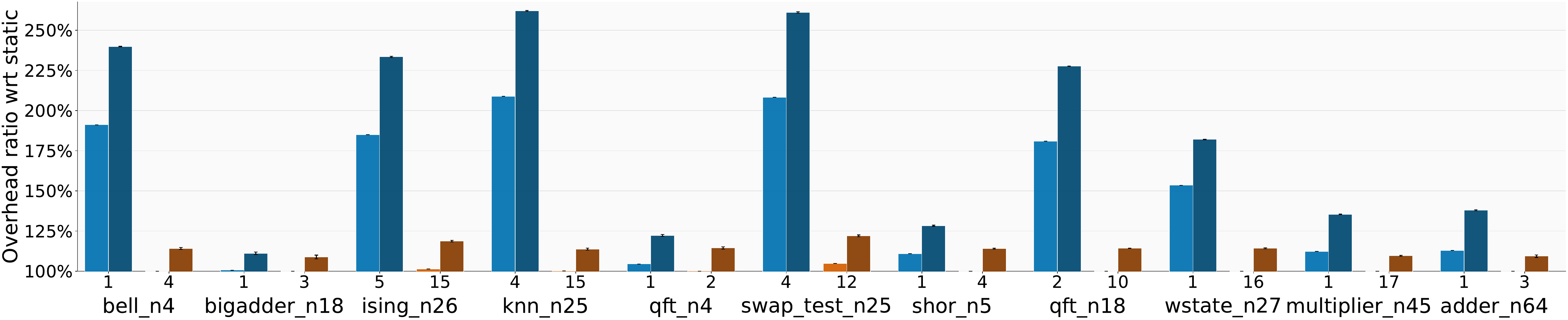}
    \caption{Cultivation at 1e-3 physical error rate. $d_1 = 3, d_2 = 7, r_1 = 3, r_2 = 5$}
    \label{fig:overhead_cycles_cultiv}
  \end{subfigure}
  \begin{subfigure}{\textwidth}
    \includegraphics[height=3.7cm,width=\columnwidth]{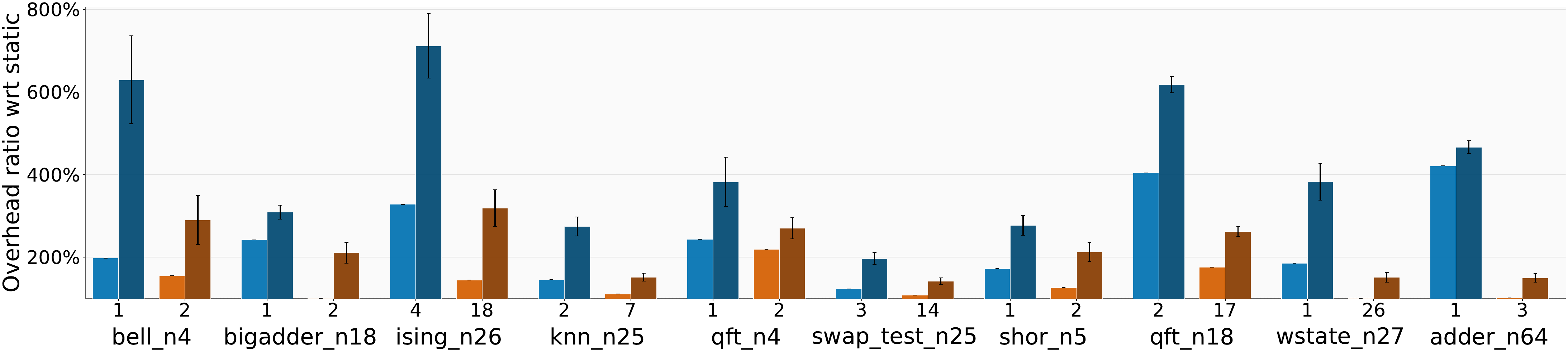}
    \caption{$R_z$ synthesis at d=3, 1e-3 physical error rate}
    \label{fig:overhead_cycles_rz}
  \end{subfigure}
  \caption{Relative cycle-time overhead when using different magic state production methods. Values are plotted as percent of each benchmark’s own static cycle count and error bars show variance in overhead across trials}
  \label{fig:overhead_cycles}
\end{figure*}

Dedicated synthesis per-angle introduces a structural difference from distillation and cultivation that has significant implications for our analysis: injection errors on an $R_z(\theta)$ gate requires a $R_z(2\theta)$ fixup, which itself requires a new magic state for an \emph{entirely different} angle (if it is not a Clifford) which may also fail. This creates a potentially cascading chain of injection failures that is unique to the Rz-synthesis paradigm. While the expected number of injections per logical $R_z(\theta)$ gate is $2$ (since each injection fails with $50\%$ probability), the latency of individual rotations exhibits nontrivial variance. Maintaining a pool of preparing units for each $\theta$, and using the first \Rz{} state produced by that pool reduces the expected stall time incurred while waiting for the magic state, but also adds to the space cost~\cite{Sethi_2025}.

\section{The Price: Non-Determinism Inflates Execution Time}
\label{sec:price}
Having defined the stochastic models, we now measure their impact on execution time. We examine each preparation mechanism independently, first across a suite of benchmarks at a fixed resource allocation, and then as a function of factory provisioning for representative circuits.

\subsection{Metrics and Methodology}
We report two primary metrics throughout this study. The first is \emph{cycle count $C$}: the number of logical cycles required to execute the circuit. The second is \emph{space-time volume}: $$V = Q_{\mathrm{total}} \times C$$ where $Q_{\mathrm{total}}$ is the total qubit count i.e. 
\begin{equation*}
\begin{aligned}
Q_{\mathrm{total}} =\;& \#\text{logical circuit qubits}\ +\\
&\left(\#\text{qubits per magic state unit} \times \#\text{production units}\right)
\end{aligned}
\end{equation*}
Space-time volume captures the full cost of execution: adding factories reduces cycle count but increases $Q_{\mathrm{total}}$, so the product $V$ reflects the true resource expenditure and reveals whether additional factories are worth their qubit cost. This metric is central to our analysis because it captures the fundamental tension that non-determinism creates: errors increase $C$ (the price) but may reduce the factory count (reducing $Q_{\mathrm{total}}$) needed to minimize $V$ (the payoff).
To isolate the contribution of each error source, we execute benchmarks under four configurations:
\begin{itemize}
    \item \textbf{Mode~A} (baseline): Both errors are disabled i.e. magic states production and injections always succeed deterministically. This yields the ideal cycle count $C_{\mathrm{ideal}}$.
    \item \textbf{Mode~B} (factory errors only): Magic state production is stochastic in nature (i.e., we model aborts and variable preparation times), but injections always succeed.
    \item \textbf{Mode~C} (injection errors only): Magic state production is deterministic, but each injection fails with $50\%$ probability, inserting fixups into the circuit.
    \item \textbf{Mode~D} (both errors): The realistic scenario, which combines both stochastic channels. This represents the realistic operating conditions for a fault-tolerant system. Modes B and C allow us to isolate factory-side effects from injection-side non-determinism.
\end{itemize}

For Clifford+T circuits, we define \emph{naive} production unit (factory) count $F_{\mathrm{naive}}$ as the number of production units required to match the circuit's non-Clifford gate demand under deterministic (error-free) production:
\begin{equation*}
    F_{\mathrm{naive}} = \left\lceil \gamma_{\mathrm{naive}} \;\times\; T_{\mathrm{prod}} \right\rceil
    \label{eq:naive_f}
\end{equation*}
where we compute ${\gamma}_\mathrm{naive}$ either as 1) the maximum per-layer non-Clifford gate count, ${\gamma}_\mathrm{peak}$ or 2) the mean per-layer non-Clifford count, ${\gamma}_\mathrm{avg}$ (computed from the deterministic schedule). Both of these choices are studied by prior work~\cite{qre1,qre2} and represent the two provisioning heuristics: allocate for worst-case peak demand, or allocate for average sustained demand, respectively. $T_{\mathrm{prod}}$ is the mechanism's expected production time per state. Our goal is to determine how far the true space-time-optimal provisioning point $F^*$ deviates from $F_{\text{naive}}$ under realistic execution.

\begin{figure*}[ht]
    \centering
    \begin{subfigure}{0.325\textwidth}
    \centering
        \includegraphics[width=6cm,height=3.5cm]{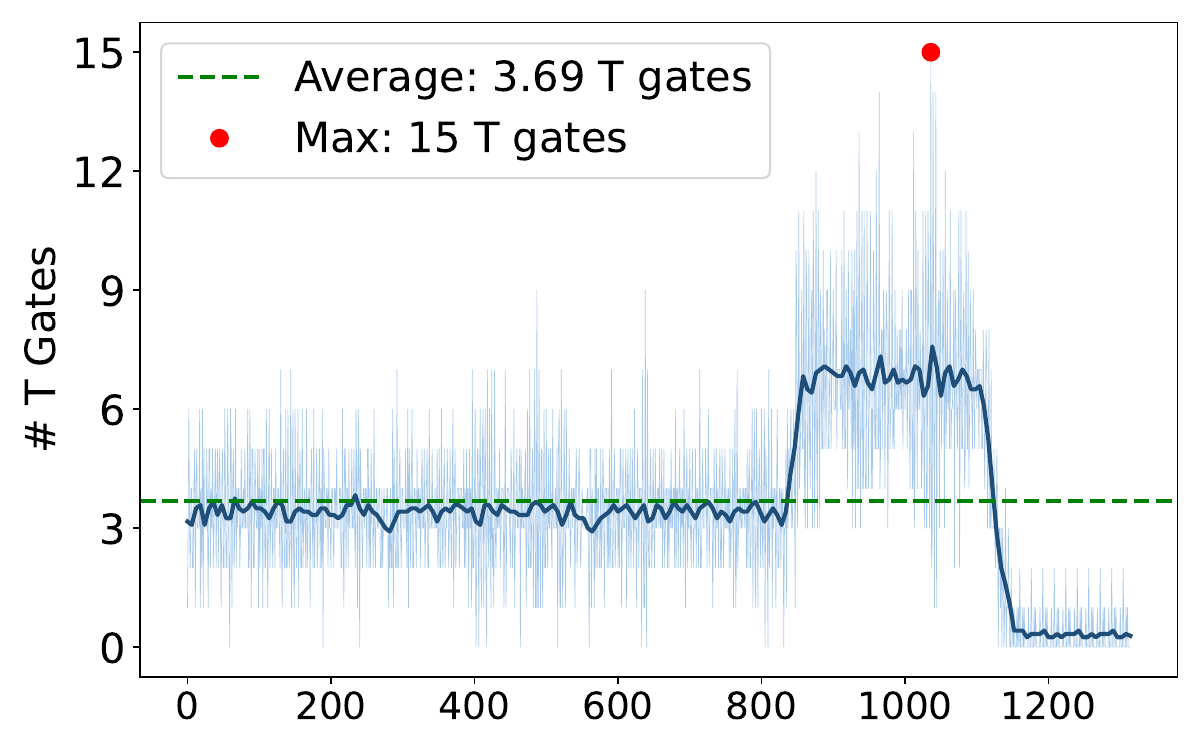}
        \label{fig:knn_n25_characterization}
    \end{subfigure}
    \begin{subfigure}{0.325\textwidth}
    \centering
        \includegraphics[width=6cm,height=3.5cm]{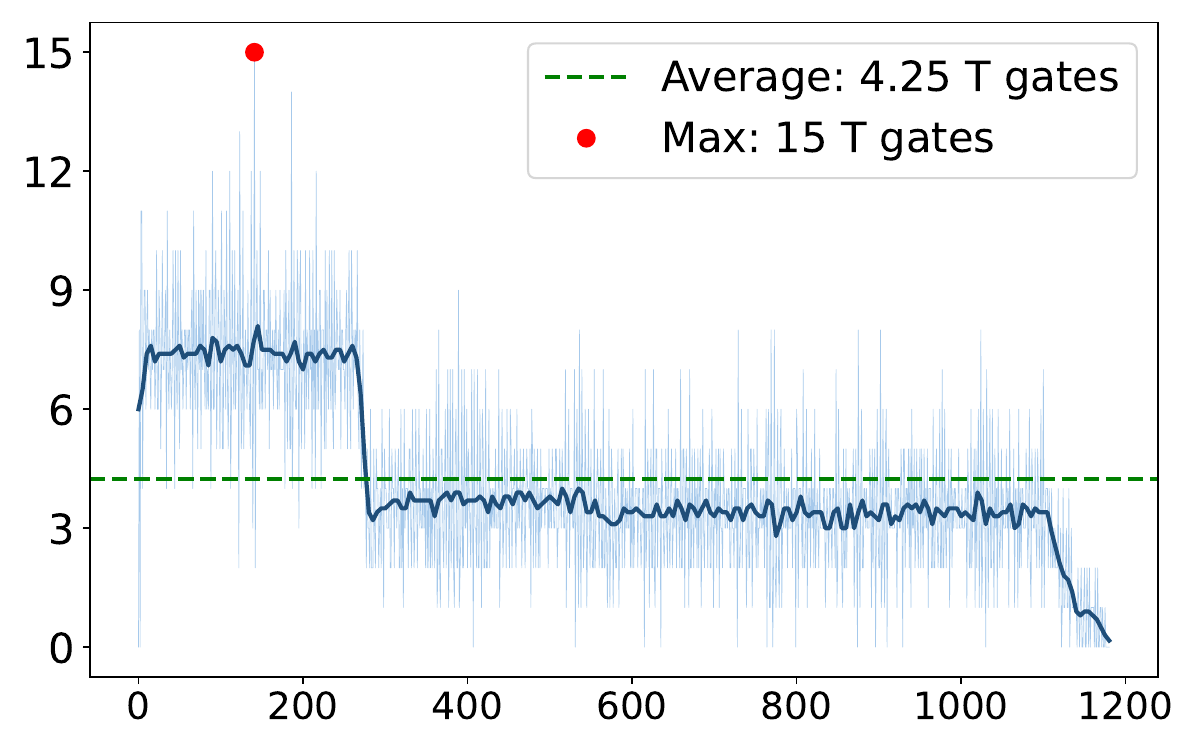}
        \label{fig:ising_n26_characterization}
    \end{subfigure}
    \begin{subfigure}{0.325\textwidth}
    \centering
        \includegraphics[width=6cm,height=3.5cm]{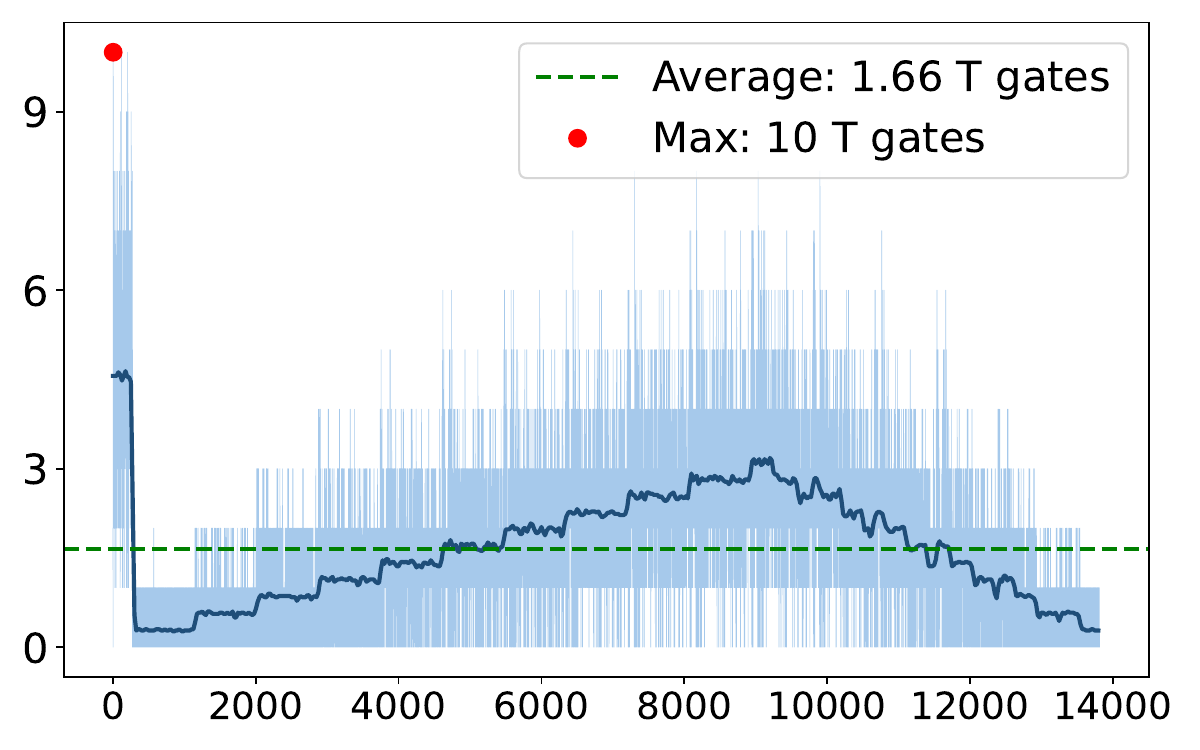}
        \label{fig:qft_n18_characterization}
    \end{subfigure}\\
    \vspace{-10pt}
    \begin{subfigure}{0.32\textwidth}
    \centering
        \includegraphics[width=5.5cm,height=3.5cm]{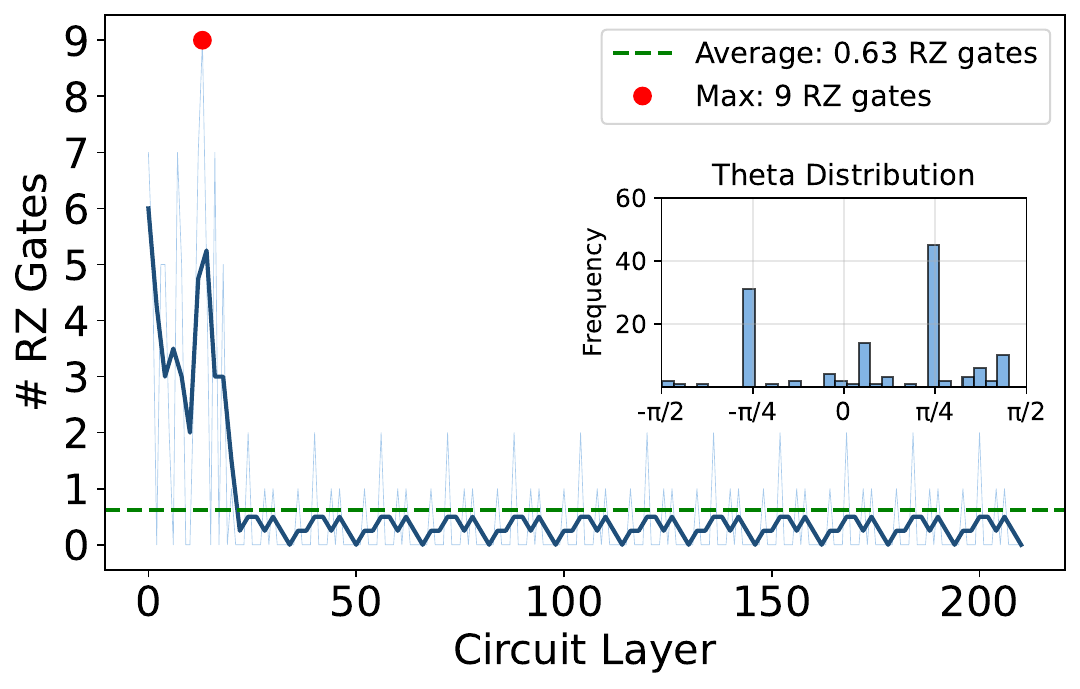}
        \caption{knn\_n25}
        \label{fig:knn_n25_rz_characterization}
    \end{subfigure}
    \begin{subfigure}{0.32\textwidth}
    \centering
        \includegraphics[width=5.5cm,height=3.5cm]{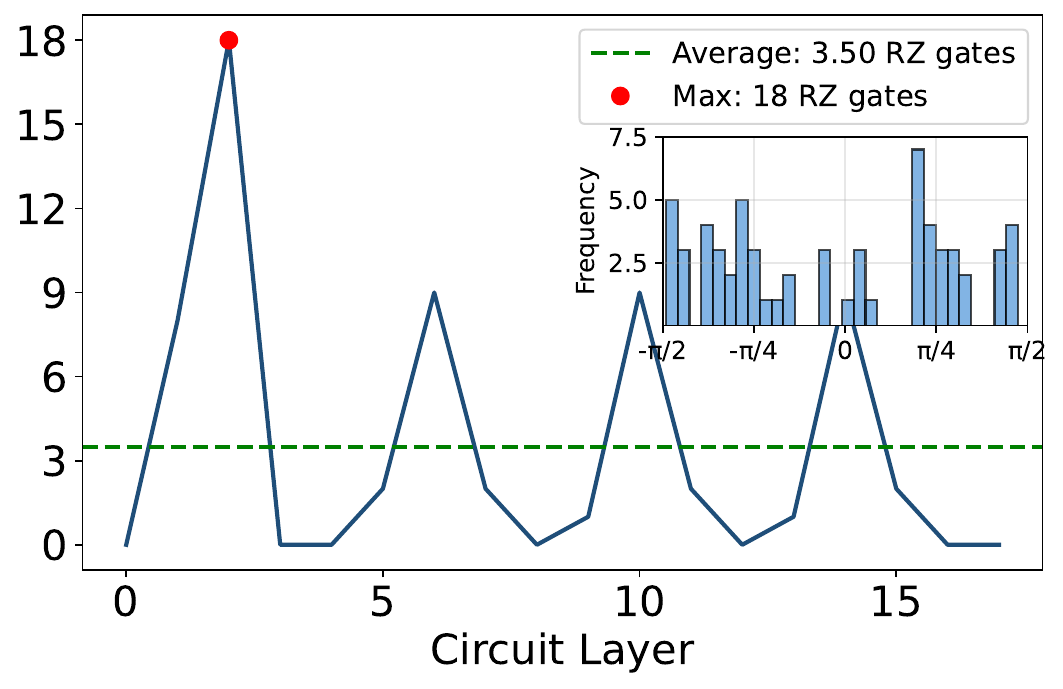}
        \caption{ising\_n26}
        \label{fig:ising_n26_rz_characterization}
    \end{subfigure}
    \begin{subfigure}{0.32\textwidth}
    \centering
        \includegraphics[width=5.5cm,height=3.5cm]{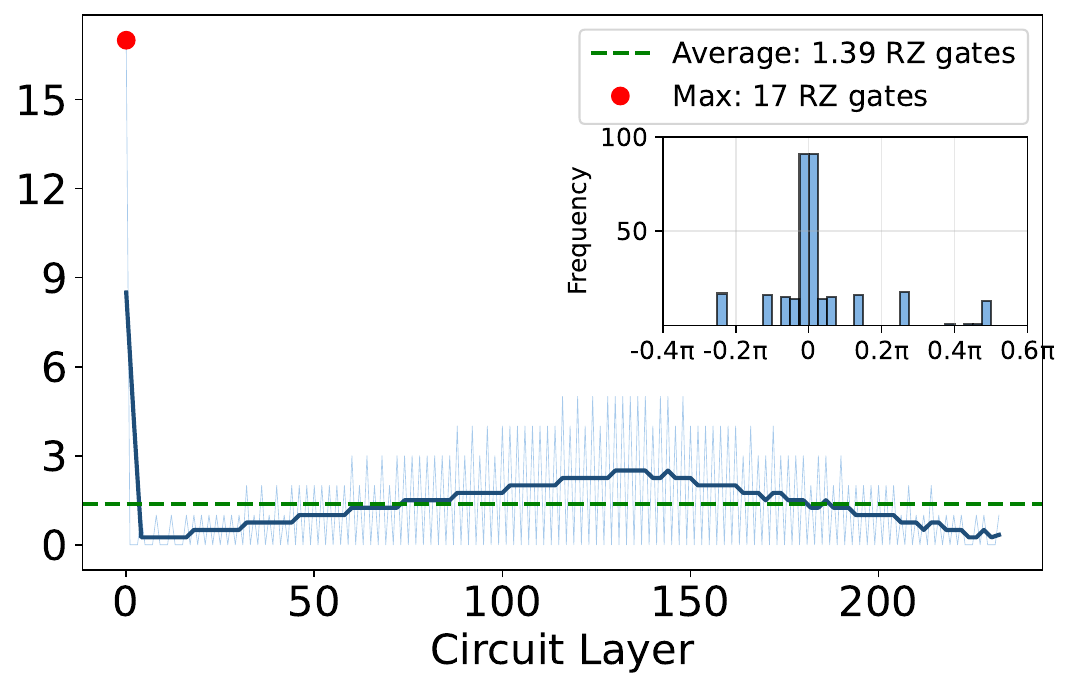}
        \caption{qft\_n18}
        \label{fig:qft_n18_rz_characterization}
    \end{subfigure}
    \caption{Static non-Clifford profiles of representative benchmarks}
    \label{fig:static_profiles}
\end{figure*}

\subsubsection{Distillation}
We evaluate the (15-to-1)$_{7,3,3}$ distillation protocol~\cite{Litinski_2019_distillation} using the configuration: physical error rate $p = 10^{-4}$, production time $T_{\text{prod}}= 18$ cycles and abort rate $\approx 0.15\%$, with factories staggered at optimal intervals.

\subsubsection{Cultivation}
We use the configuration $d_1{=}3$, $d_2{=}7$, $r_1{=}3$, $r_2{=}5$ at physical error rate $p{=}10^{-3}$, with factories feeding a global buffer (of the same size as the number of factories).

\subsubsection{Direct \Rz{} Synthesis}
We evaluate \Rz{} synthesis at code distance $d{=}3$ with physical error rate $p{=}10^{-3}$, sweeping over number of production units.

\subsection{Observations}
Figure~\ref{fig:overhead_cycles} presents the cycle-time overhead ratio $\frac{C}{C_\text{static}}$ for benchmark circuits from QASMBench~\cite{li2022qasmbenchlowlevelqasmbenchmark}, evaluated at both $F_{\mathrm{naive}}$ provisioning levels ($\gamma_\text{avg}$ and $\gamma_\text{peak}$). Two trends are immediately apparent. First, non-determinism always inflates execution time; no benchmark escapes overhead-free under Mode D. Second, the overhead is consistently lower in the $\gamma_\text{peak}$ allocation regime than in $\gamma_\text{avg}$ i.e. additional factories absorb stochastic supply shortfalls, confirming that provisioning level mediates the severity of the price. However, the gap between the two regimes varies substantially across benchmarks and mechanisms, indicating that a single provisioning heuristic cannot entirely capture circuit-specific behavior.

\begin{figure*}[ht]
    \centering
    \begin{subfigure}{\textwidth}
    \centering
        \includegraphics[width=\linewidth]{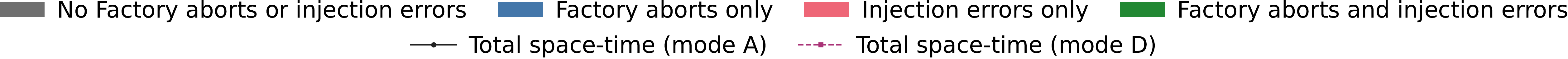}
    \end{subfigure}\\
    \vspace{2pt}
    \begin{subfigure}{0.31\textwidth}
    \centering
        \includegraphics[width=6cm,height=3.8cm]{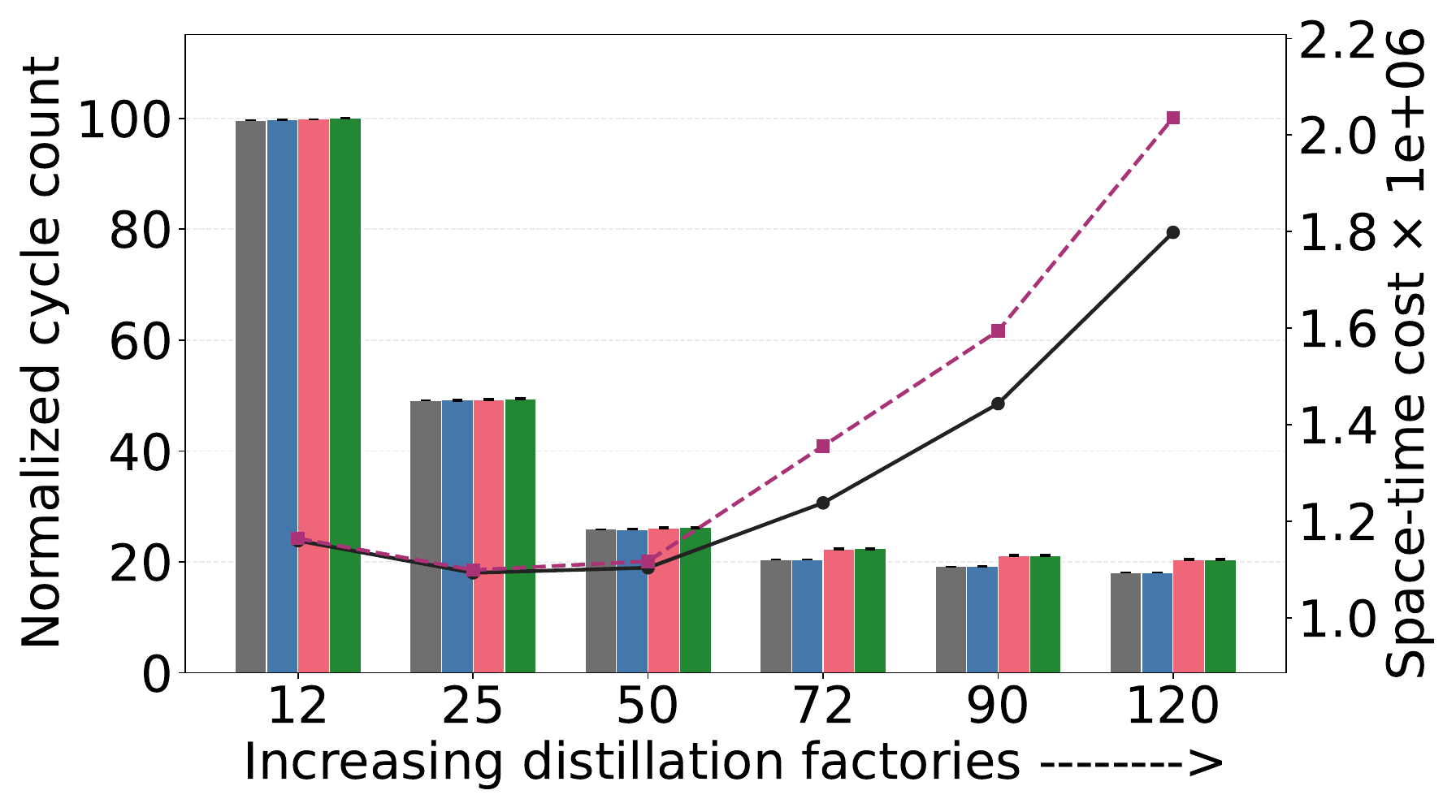}
        \caption{knn\_n25 under 15-to-1 distillation}
        \label{fig:knn_distil_absolute}
    \end{subfigure}
    \hfill
    \begin{subfigure}{0.31\textwidth}
    \centering
        \includegraphics[width=6cm,height=3.8cm]{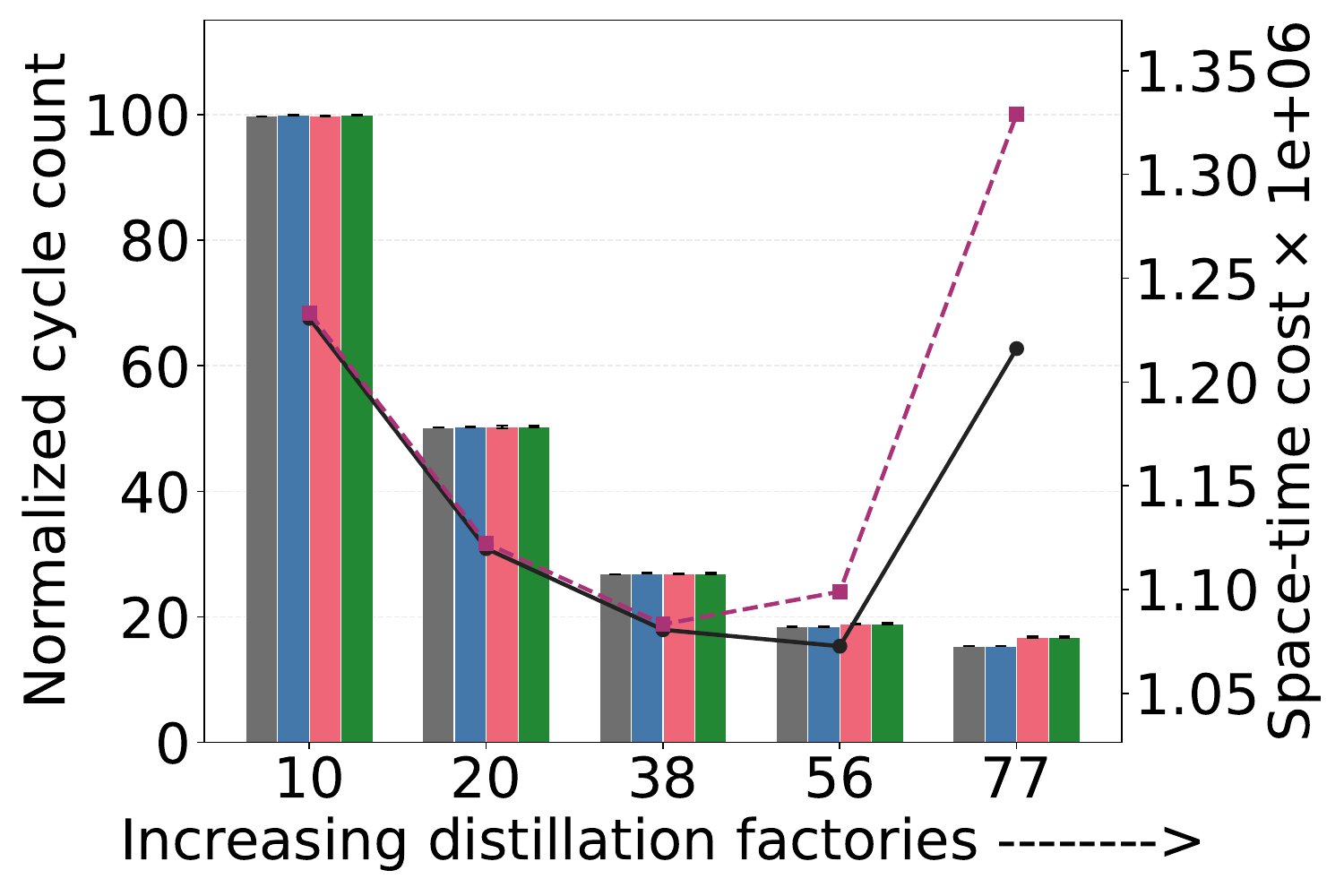}
        \caption{ising\_n26 under 15-to-1 distillation}
        \label{fig:ising_distil_absolute}
    \end{subfigure}
    \hfill
    \begin{subfigure}{0.31\textwidth}
    \centering
        \includegraphics[width=6cm,height=3.8cm]{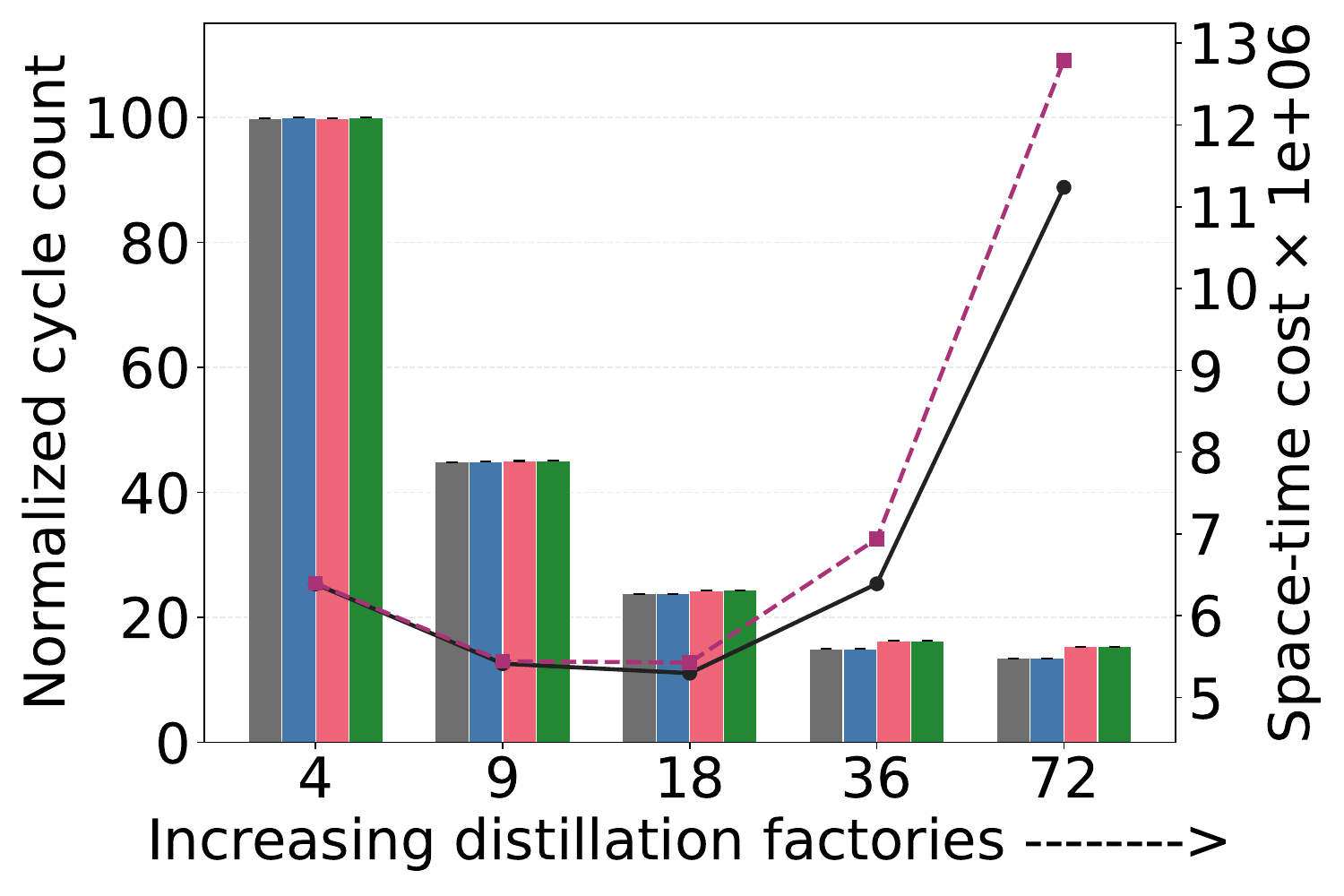}
        \caption{qft\_n18 under 15-to-1 distillation}
        \label{fig:qft_distil_absolute}
    \end{subfigure}
    \hfill
    \begin{subfigure}{0.31\textwidth}
    \centering
        \includegraphics[width=6cm,height=3.8cm]{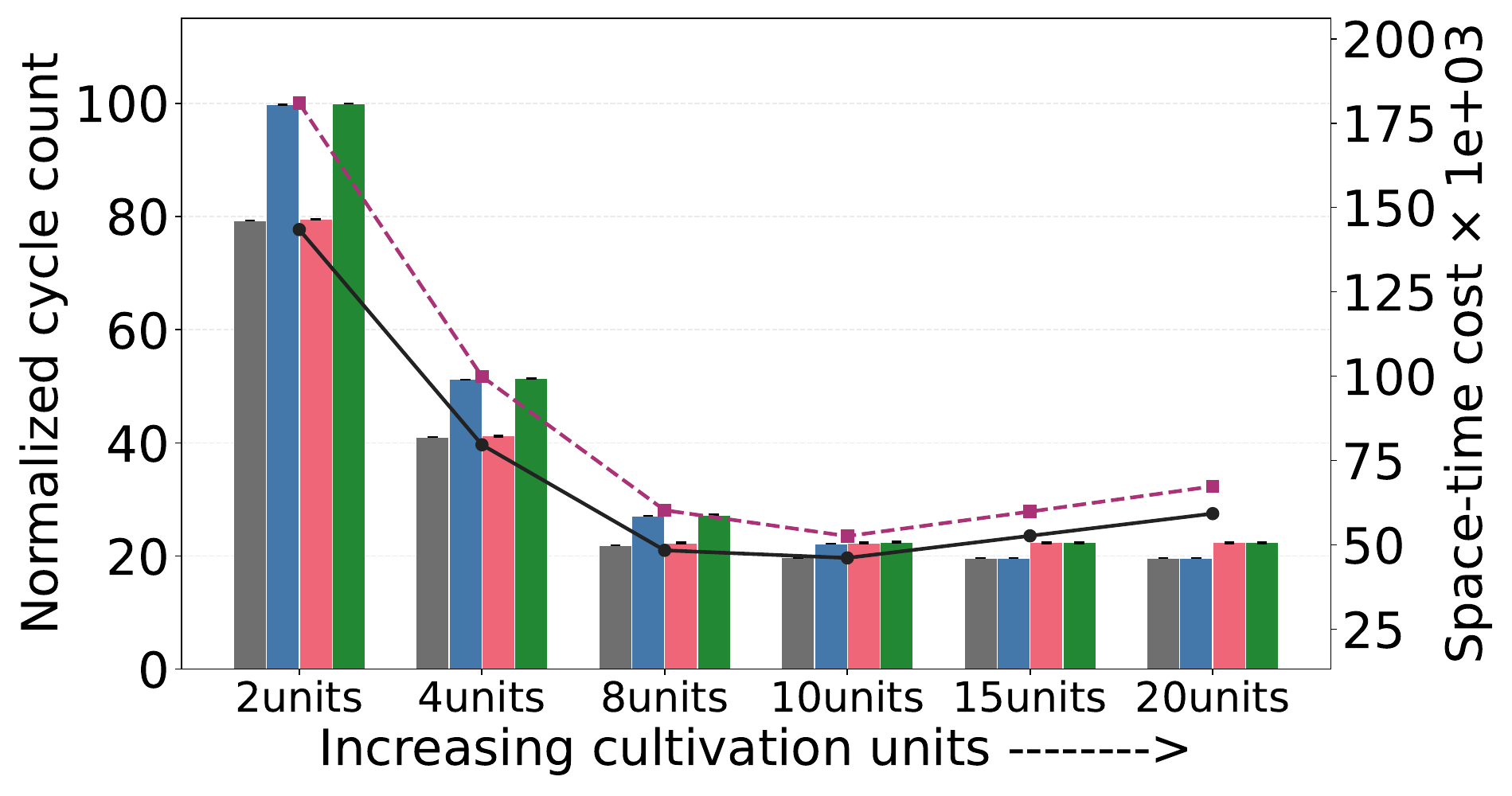}
        \caption{knn\_n25 under cultivation}
        \label{fig:knn_cultiv_absolute}
    \end{subfigure}
    \hfill
    \begin{subfigure}{0.31\textwidth}
    \centering
        \includegraphics[width=6cm,height=3.8cm]{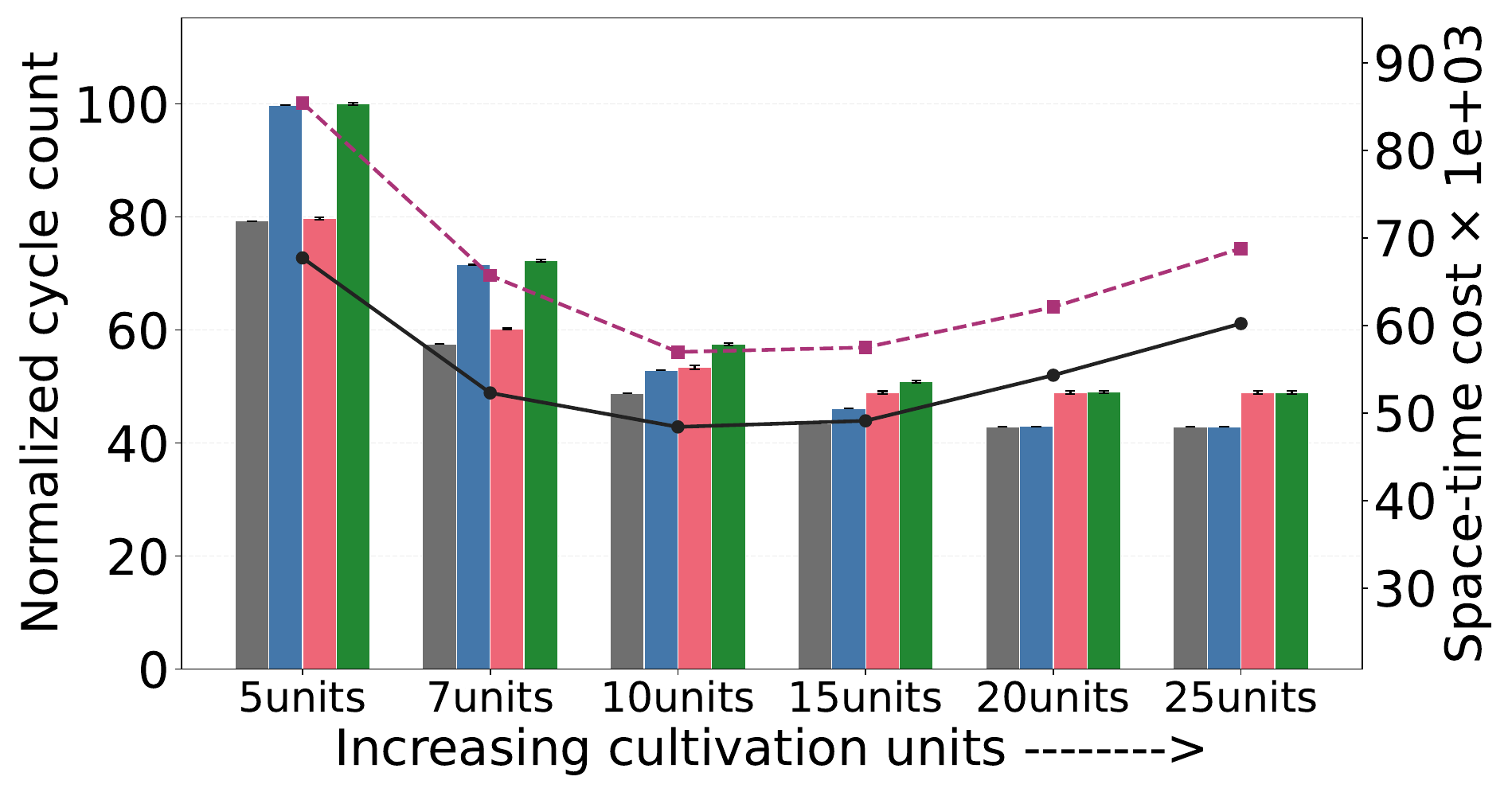}
        \caption{ising\_n26 under cultivation}
        \label{fig:ising_cultiv_absolute}
    \end{subfigure}
    \hfill
    \begin{subfigure}{0.31\textwidth}
    \centering
        \includegraphics[width=6cm,height=3.8cm]{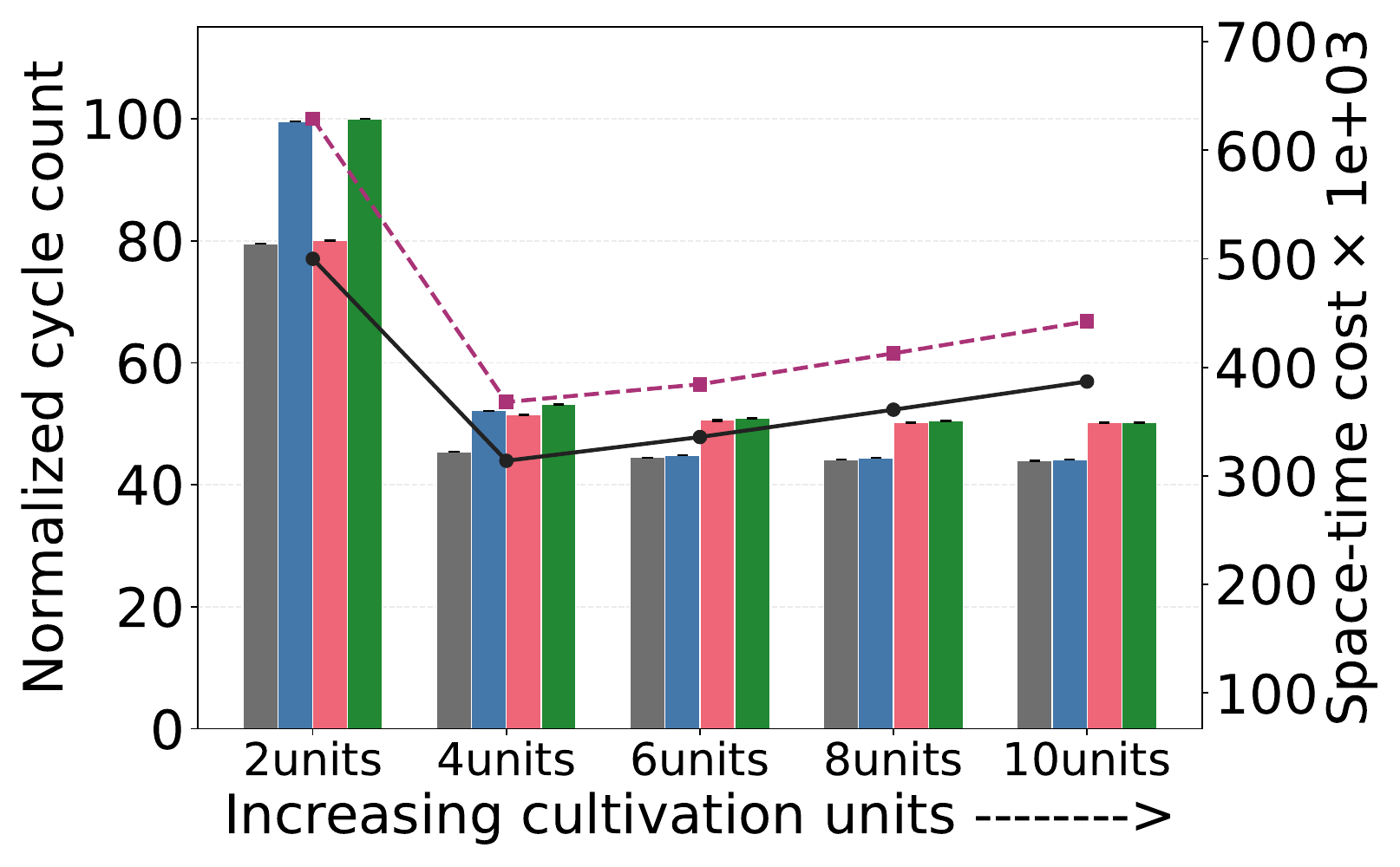}
        \caption{qft\_n18 under cultivation}
        \label{fig:qft_cultiv_absolute}
    \end{subfigure}
    \hfill
    \begin{subfigure}{0.31\textwidth}
    \centering
        \includegraphics[width=6cm,height=3.8cm]{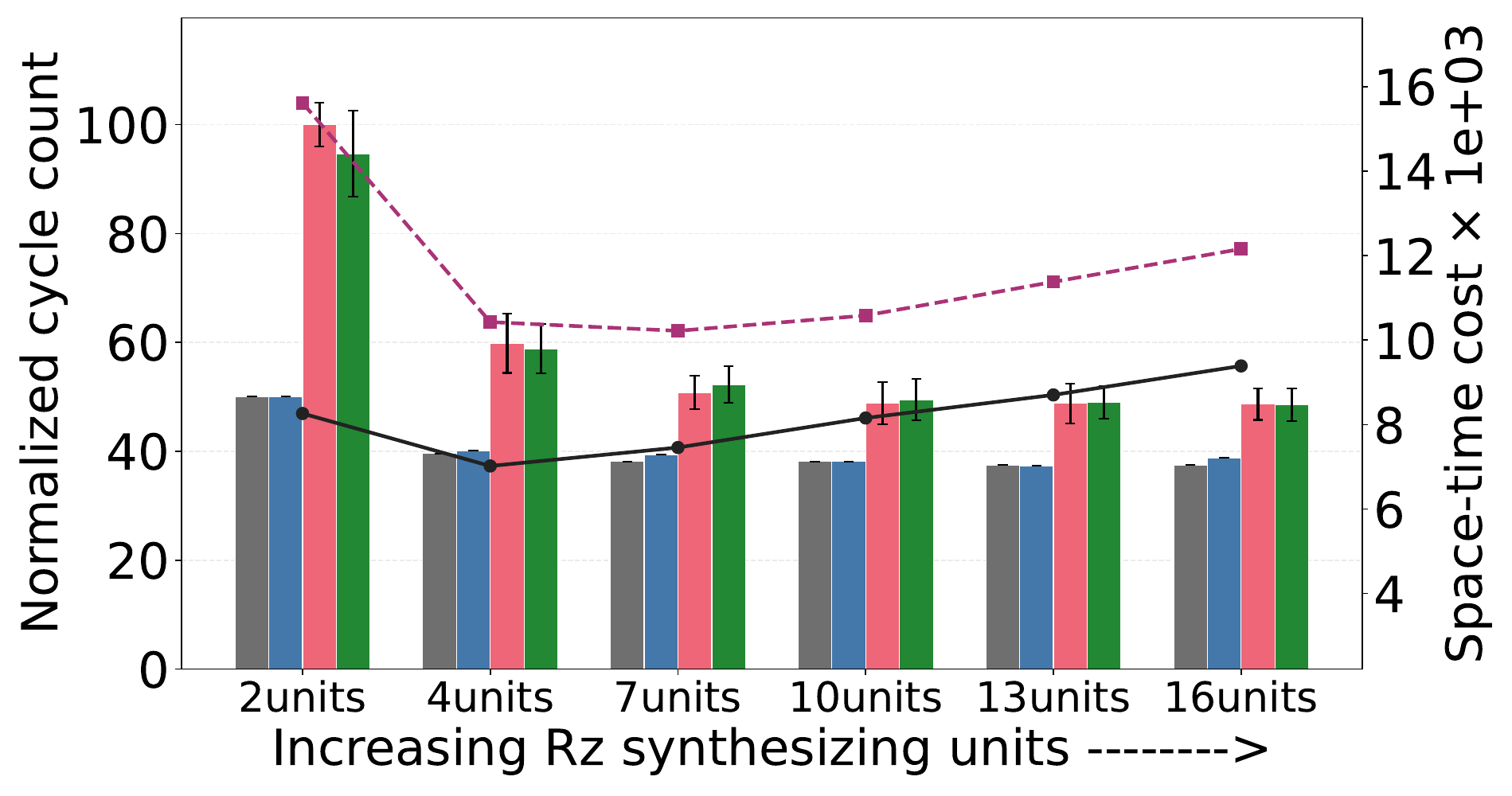}
        \caption{knn\_n25 under $R_z$ synthesis}
        \label{fig:knn_rz_absolute}
    \end{subfigure}
    \hfill
    \begin{subfigure}{0.31\textwidth}
    \centering
        \includegraphics[width=6cm,height=3.8cm]{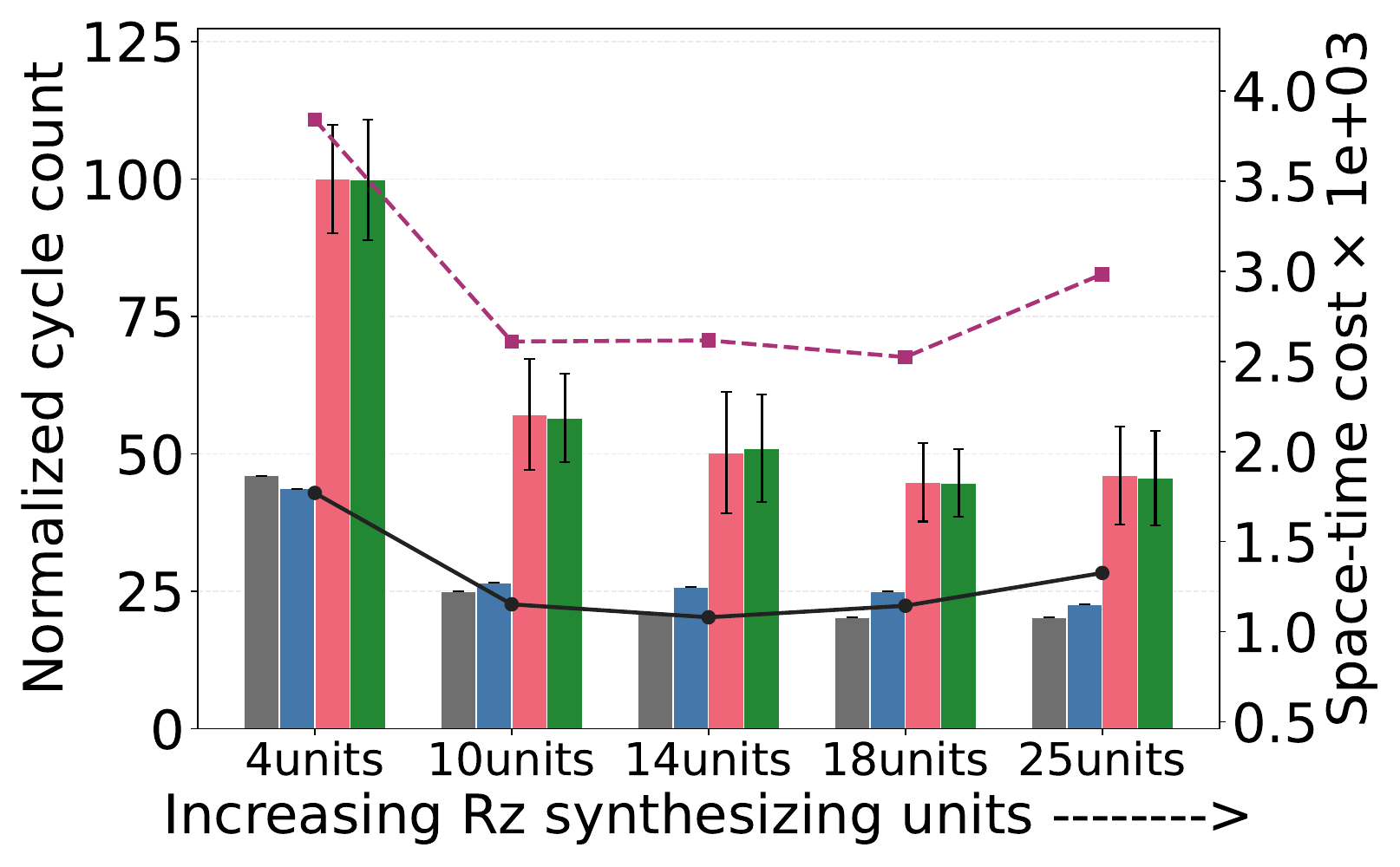}
        \caption{ising\_n26 under $R_z$ synthesis}
        \label{fig:ising_rz_absolute}
    \end{subfigure}
    \hfill
    \begin{subfigure}{0.31\textwidth}
    \centering
        \includegraphics[width=6cm,height=3.8cm]{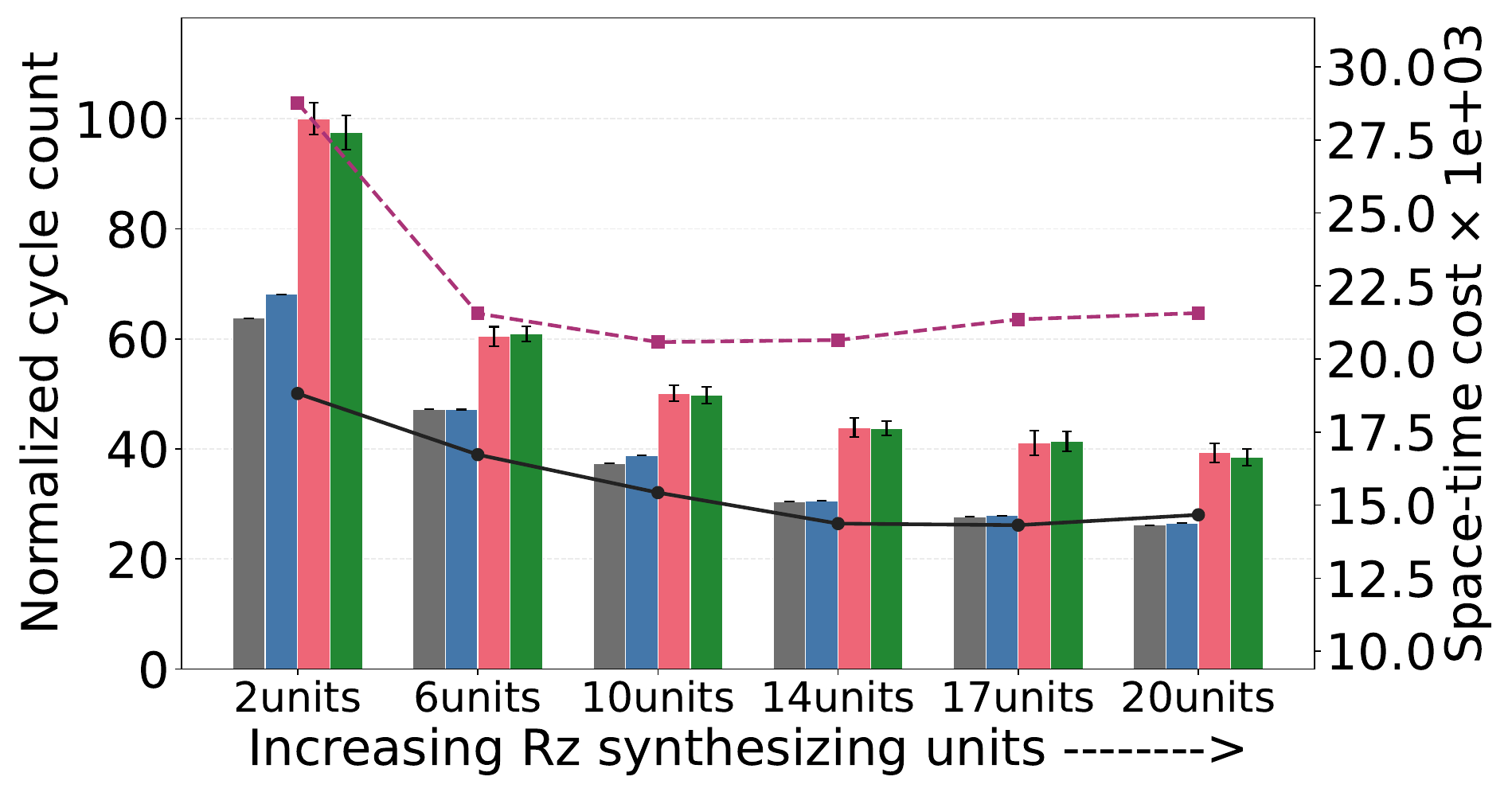}
        \caption{qft\_n18 under $R_z$ synthesis}
        \label{fig:qft_rz_absolute}
    \end{subfigure}
    \caption{Cycle count and space-time cost vs. \#magic state production units under different production methods. The space-time cost has a convex structure, and a clearly identifiable minimum. In the low factory regime, increasing \#production units leads to major reduction in cycle counts. However, on increasing \#production units indiscriminately, we get almost no benefit in cycle count for high increase in space-time cost}
    \label{fig:cycleandspacetime}
\end{figure*}

Figure~\ref{fig:cycleandspacetime} shows how the cycle overhead and space-time cost evolve as the number of production units, $F$, increases for representative benchmarks (Figure~\ref{fig:static_profiles} shows their static profiles) under each mechanism.
At low~$F$, throughput is insufficient to keep pace with demand even without errors, so all modes exhibit high cycle counts. As $F$ increases, the error-free cycle count (Mode~A) drops and \emph{eventually plateaus} at the point where factories are no longer the bottleneck and the circuit's inherent depth limits further improvement. The stochastic modes (B, C, D) follow the same shape but are shifted upward by their respective overheads. Beyond the plateau, $C$ is flat but $Q_\text{total}$ continues to grow, so $V$ increases linearly.
The critical observation is the tension between cycle count and space-time volume: each additional production unit reduces the gap between deterministic and realistic execution, but adds qubits to $Q_\text{total}$. The space-time cost therefore exhibits a convex structure with a clearly identifiable minimum: the optimal provisioning point $F^*$ that we seek.

Factory-level failures reduce effective throughput. When a production unit fails to deliver a state on schedule, any dependent non-Clifford gate must stall until the next successful round from any available unit. The severity scales inversely with provisioning such that with ample factories, other units absorb the shortfall transparently. With marginal provisioning, a single abort can stall the critical path and result into downstream delays. Injection errors have a qualitatively different effect as they reshape demand by inserting fixup operations into the circuit DAG. Each fixup extends the path length by at least one cycle, and this may also reshuffle the execution order via priority recomputation. The net impact depends on whether the fixup lands on the critical path, the DAG's frontier width at that point, and current factory availability. For \Rz{} synthesis, the effect compounds, as it may trigger cascading resource allocations. This behavior is evident in Figure~\ref{fig:overhead_cycles_rz}, where \Rz{} synthesis overheads reach up to 8$\times$ under minimal provisioning, far exceeding the other two mechanisms.

\section{The Payoff: Non-Determinism Reshapes Resource Demand}
\label{sec:payoff}
The previous section established that non-determinism always inflates execution time. We now show that the same stochastic effects simultaneously provide another benefit: they \emph{reduce} the peak per-cycle demand for magic states below what deterministic analysis predicts. This demand smoothing has a direct architectural consequence where fewer production units than $F_{\text{naive}}$ can sustain execution without excessive stalling, potentially lowering the spacetime-optimal factory count.

\subsection{Deterministic vs. Realized Demand}
Under deterministic (error-free) execution, the per-cycle demand for non-Clifford gates is fully determined by the circuit's DAG structure. Each gate is scheduled at the earliest cycle permitted by its dependencies, producing a fixed demand profile. Under realistic (stochastic) execution, two mechanisms redistribute this demand:
\begin{enumerate}
    \item Injection-error fixups insert additional operations that de-synchronize gates originally scheduled in the same cycle, spreading them across multiple cycles.
    \item Factory stalls force some gates to wait, dispersing their demand (and of their dependents) into later cycles.
\end{enumerate}

\begin{figure}[t]
    \centering
\includegraphics[width=\linewidth,height=4.5cm]{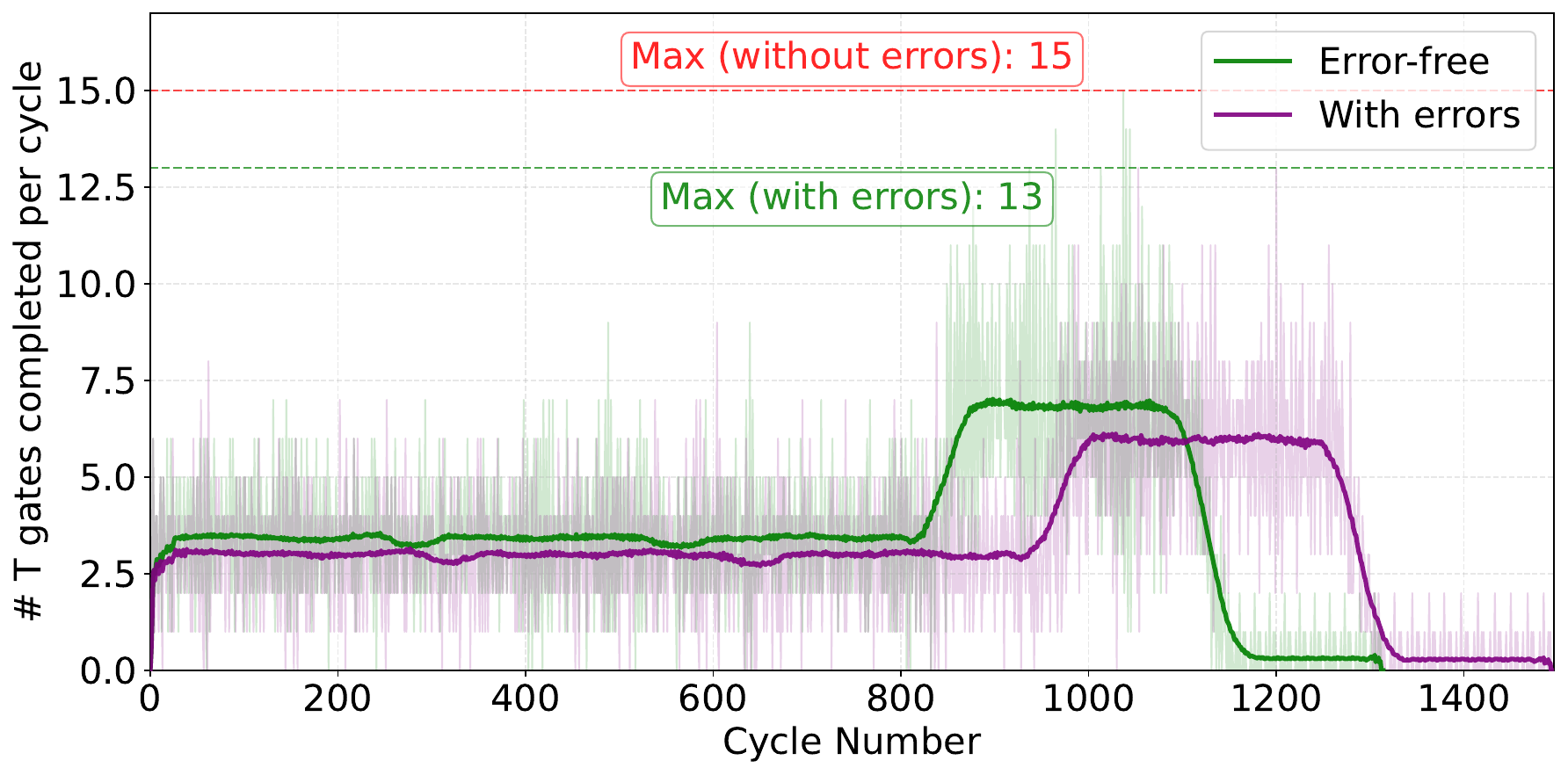}
    \caption{For knn\_n25, the deterministic trace (green) exhibits higher spikes corresponding to layers with high T-gate parallelism. In the stochastic trace (purple), the spikes are attenuated and demand is redistributed into following cycles}
    \label{fig:det_vs_non-det}
\end{figure}

The combined effect is a \emph{temporal smoothing} of the demand profile. Figure~\ref{fig:det_vs_non-det} illustrates this for knn\_n25 under distillation where the deterministic trace has sharp spikes reaching up to 15 T gates per cycle, while the stochastic trace attenuates these peaks to a maximum of 13 resulting in a a 13.3\% reduction in peak demand, while redistributing the load across additional cycles. The total number of non-Clifford gates consumed is at least as large as in the deterministic case (strictly larger for \Rz{} synthesis, where fixups introduce new rotation angles), but these gates are spread over more cycles, possibly reducing the instantaneous throughput that factories must sustain.

\subsection{Plateauing and Diminishing Returns}
The demand smoothing has a direct consequence for provisioning. If stochastic execution never reaches the deterministic peak demand, then factories provisioned to handle that peak are wasteful. Concretely, if we provision $\gamma_{\mathrm{peak}} \times T_{\mathrm{prod}}$ factories, a fraction of that capacity is wasted because errors spread demand below the provisioned throughput.

\begin{figure}[b]
    \centering
    \includegraphics[width=\linewidth,height=4cm]{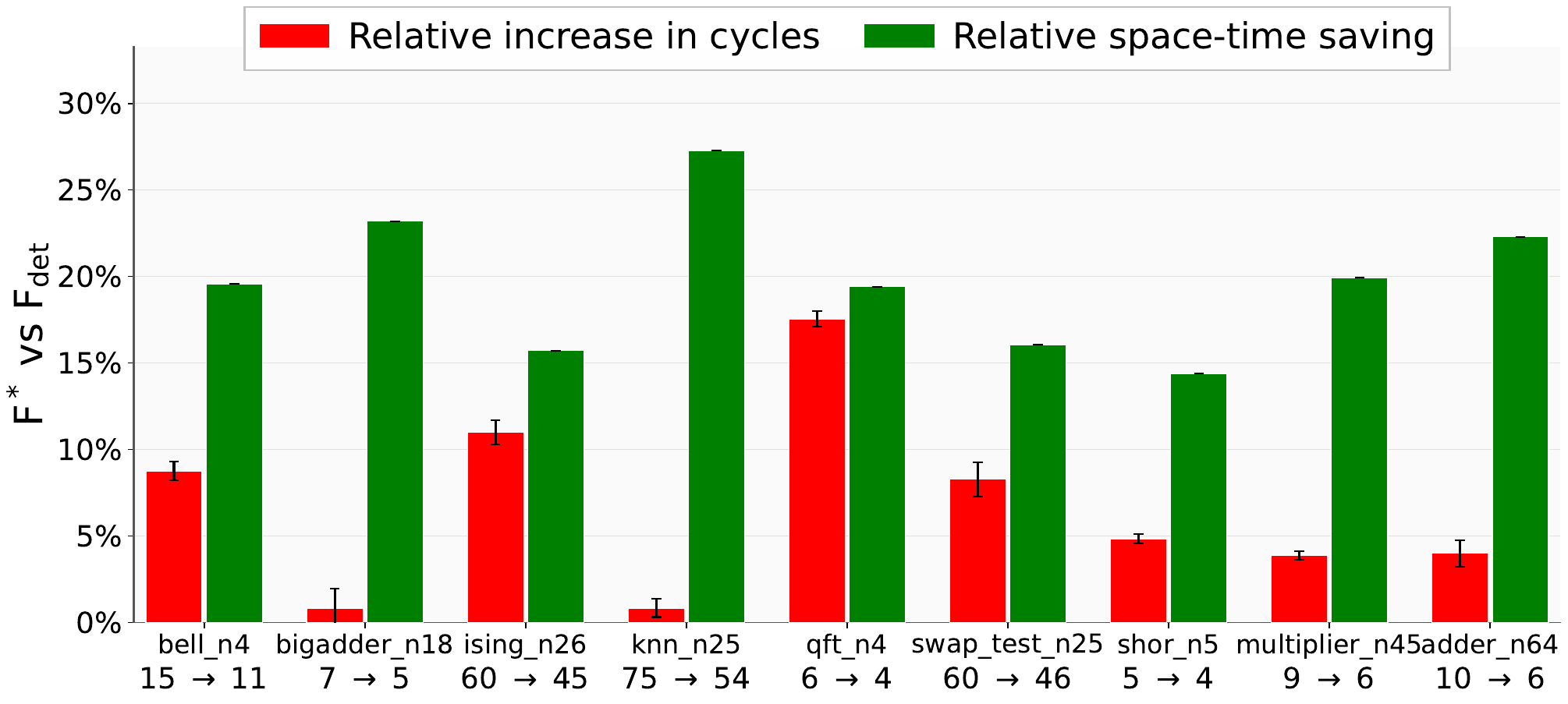}
    \caption{Relative change in cycle count and space-time cost when reducing the factory allocation (shown as $F_\text{det} \rightarrow F^*$ below each benchmark). The blue bars show the percentage increase in cycles due to fewer factories, while the amber bars show the corresponding percentage saving in space-time cost. $F^*$ achieves an overall favorable trade-off by reducing space-time cost}
    \label{fig:payoff_summary}
\end{figure}

\subsubsection{Distillation}
\begin{figure}[t]
    \centering
    \begin{subfigure}{\linewidth}
    \centering
        \includegraphics[width=\linewidth]{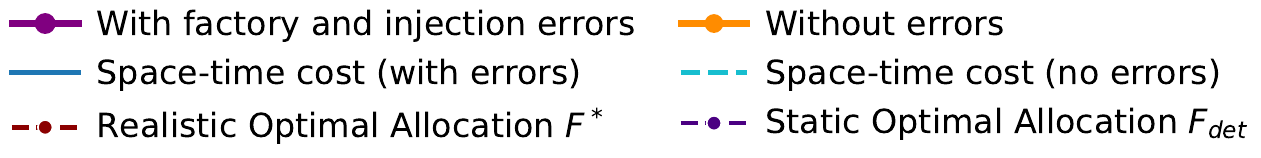}
    \end{subfigure}\\
    \begin{subfigure}{\linewidth}
        \centering
        \includegraphics[width=0.85\linewidth,height=3.9cm]{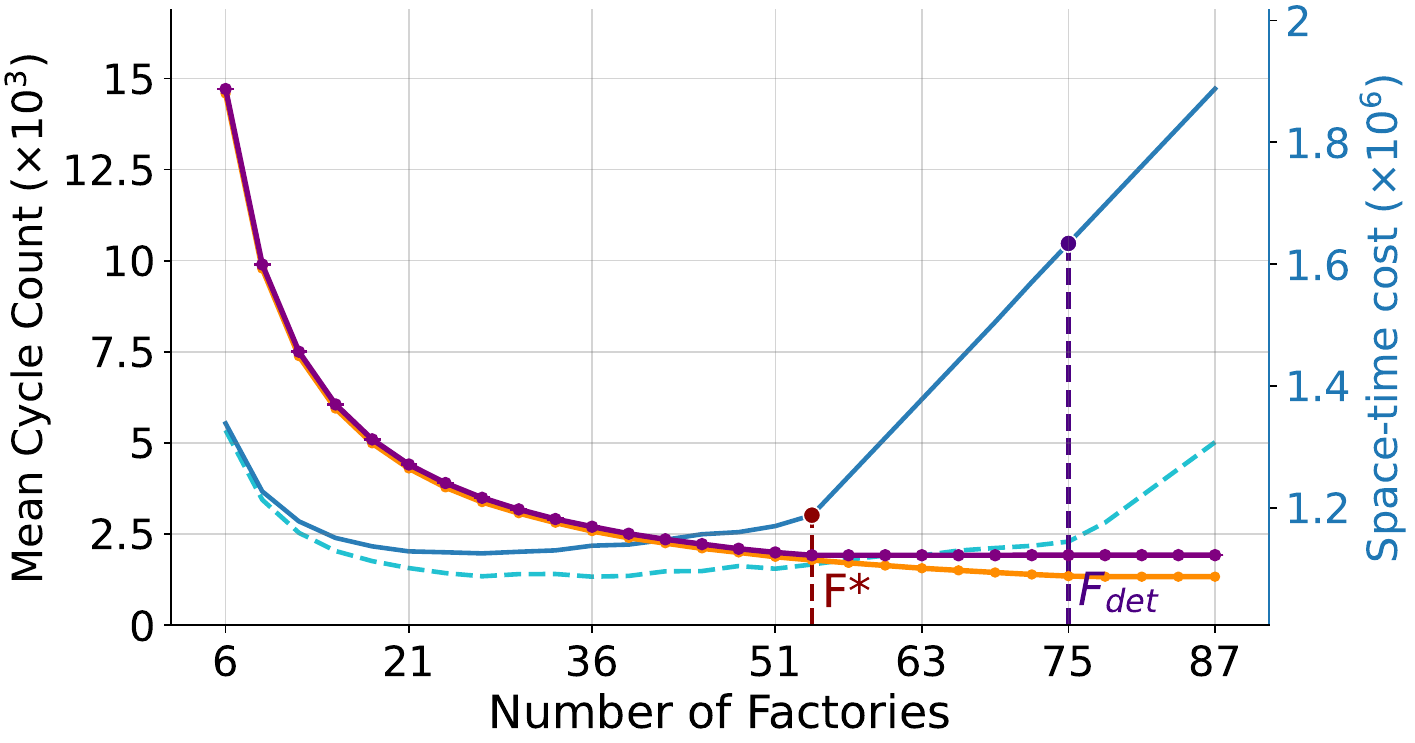}
        \caption{knn\_n25 under 15-to-1 distillation}
        \label{fig:distil_knn_n25_expdecay}
    \end{subfigure}
    \begin{subfigure}{\linewidth}
        \centering
        \includegraphics[width=0.85\linewidth,height=3.9cm]{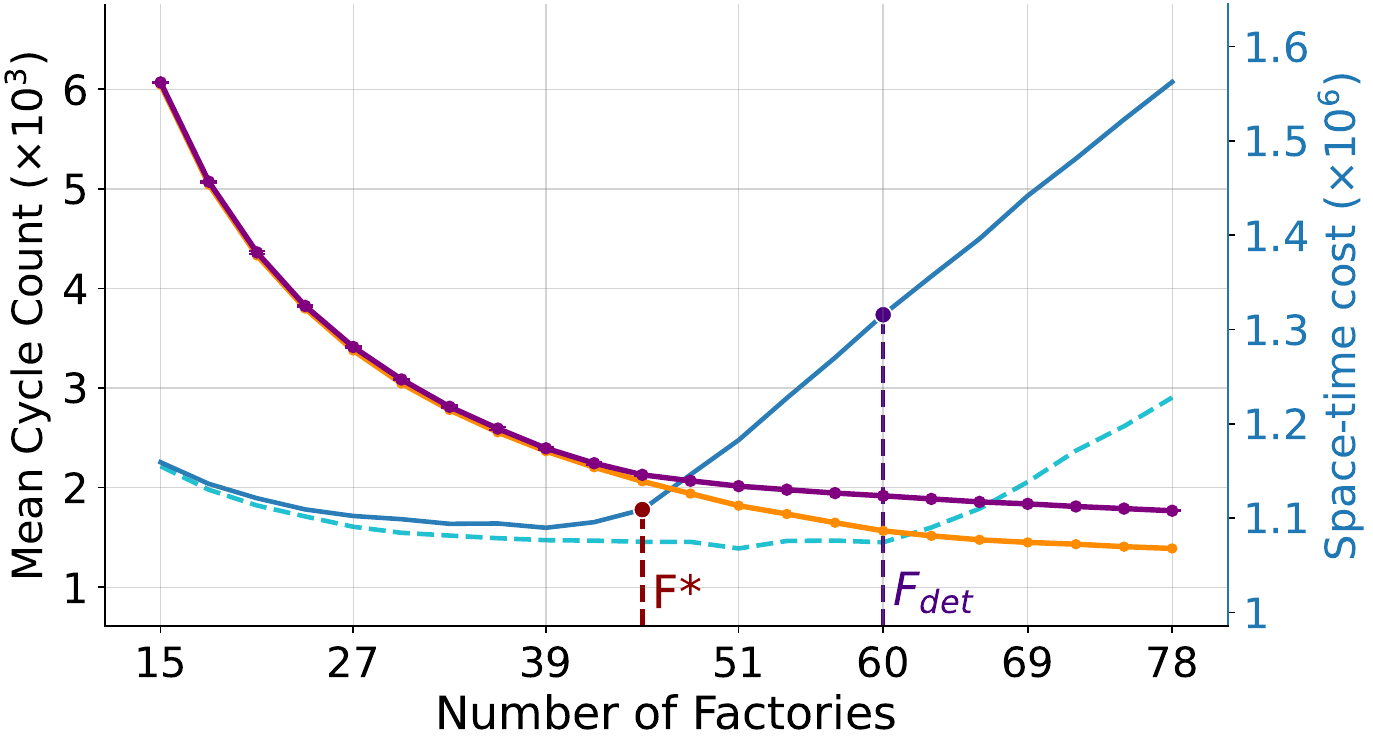}
        \caption{ising\_n26 under 15-to-1 distillation}
        \label{fig:distil_ising_n26_expdecay}
    \end{subfigure}
    \caption{The deterministic cycle count (orange) and stochastic cycle count (purple) plateau on increasing production units. The starting point of the plateau corresponds to point from which the space-time cost starts increasing almost linearly. For distillation, the stochastic curve saturates at a lower $F=F^*$.}
    \label{fig:expdecay}
\end{figure}

Figures~\ref{fig:distil_knn_n25_expdecay} and  \ref{fig:distil_ising_n26_expdecay} show the cycle count and space-time cost as a function of $F$ for three benchmarks under 15-to-1 distillation. Both the deterministic and stochastic curves plateau as $F$ increases, but the stochastic curve saturates at a strictly lower $F$. Once demand smoothing has eliminated the sharpest per-cycle spikes, additional factories provide no further cycle-count reduction. The gap between the two plateau points $F_\text{savings} = F_\text{det} - F^*$ represents factories that deterministic analysis deems necessary but that stochastic execution never simultaneously utilizes. For example, on knn\_n25 for distillation, $F^* = 54$ while $F_{\text{det}} = 75$, yielding $F_{\text{savings}} = 21$ factories. Since each factory occupies ~810 physical qubits~\cite{Litinski_2019_distillation}, 21 unnecessary factories represent over 18,000 physical qubits, which would have been wastefully allocated by deterministic analysis. This effect is seen across the benchmark suite (see Figure~\ref{fig:payoff_summary}), and the gap translates directly into qubit savings. The space-time volume captures this tradeoff in a single metric. Below the plateau, adding a factory reduces $C$ enough to offset the increase in $Q_{\text{total}}$; beyond the plateau, $C$ is almost flat but $Q_{\text{total}}$ grows linearly, so $V$ increases. As experiments show, $F^*$ under stochastic execution is strictly lower than $F_{\text{det}}$ under deterministic execution, confirming that deterministic provisioning systematically over-allocates qubit budget for distillation-based architectures.\\

\subsubsection{Cultivation and \Rz{}-synthesis}
Cultivation and \Rz{} synthesis present a different picture. The stochastic curve plateaus at roughly the same factory count as the deterministic curve; the demand smoothing does not shift $F^*$ meaningfully. These production units have a much smaller spatial footprint and much shorter average production times than distillation factories. Their high failure rates are offset by fast retries, so even under deterministic execution the circuit reaches its depth-limited plateau at a moderate factory count. Since these units are cheap and fast, the additional demand from injection-error fixups is absorbed without requiring significantly more factories than the deterministic case.

\begin{figure*}[t]
    \centering
    \begin{subfigure}{0.31\textwidth}
        \includegraphics[width=6cm,height=4cm]{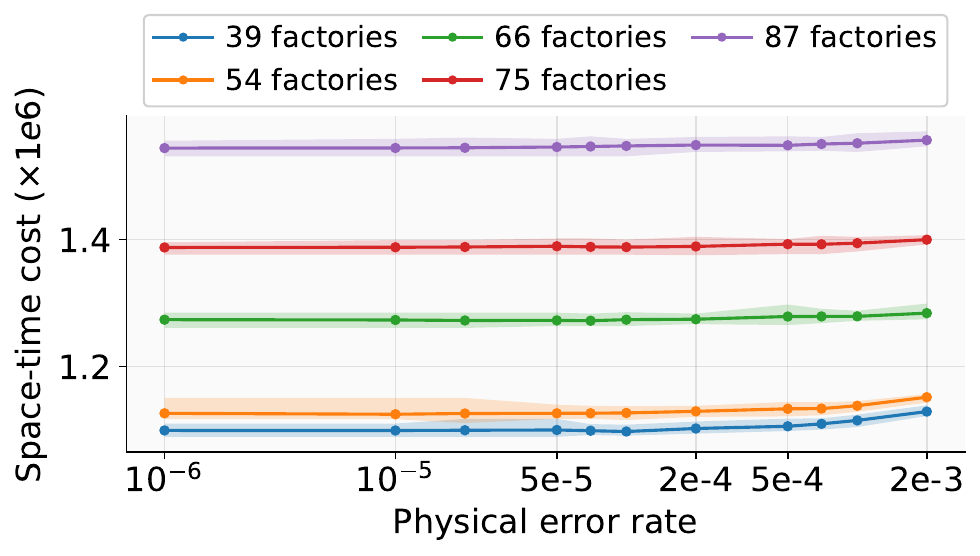}
        \caption{knn\_n25 for Distillation}
        \label{fig:distil_sensitivity}
    \end{subfigure}
    \hfill
    \begin{subfigure}{0.31\textwidth}
        \includegraphics[width=6cm,height=4cm]{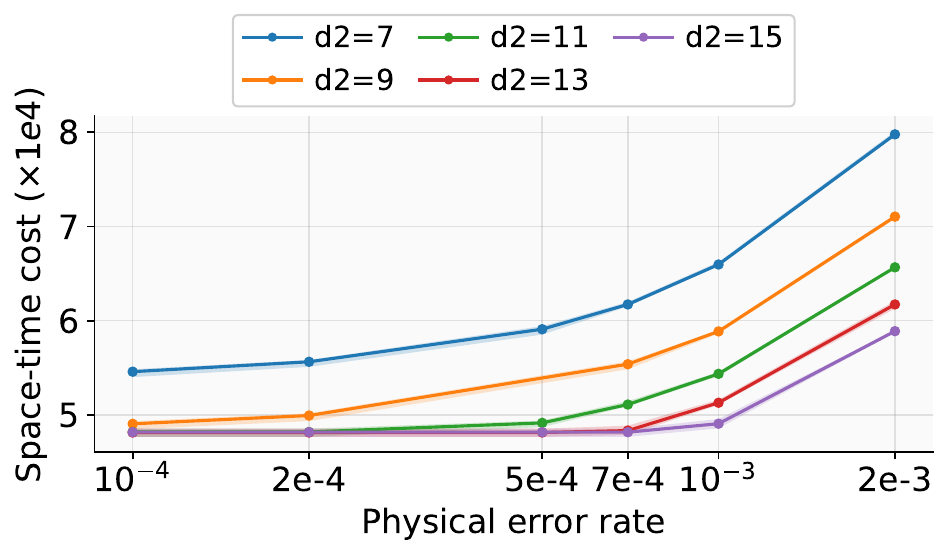}
        \caption{knn\_n25 for Cultivation}
        \label{fig:cultiv_sensitivity}
    \end{subfigure}
    \hfill
    \begin{subfigure}{0.31\textwidth}
        \includegraphics[width=6cm,height=4cm]{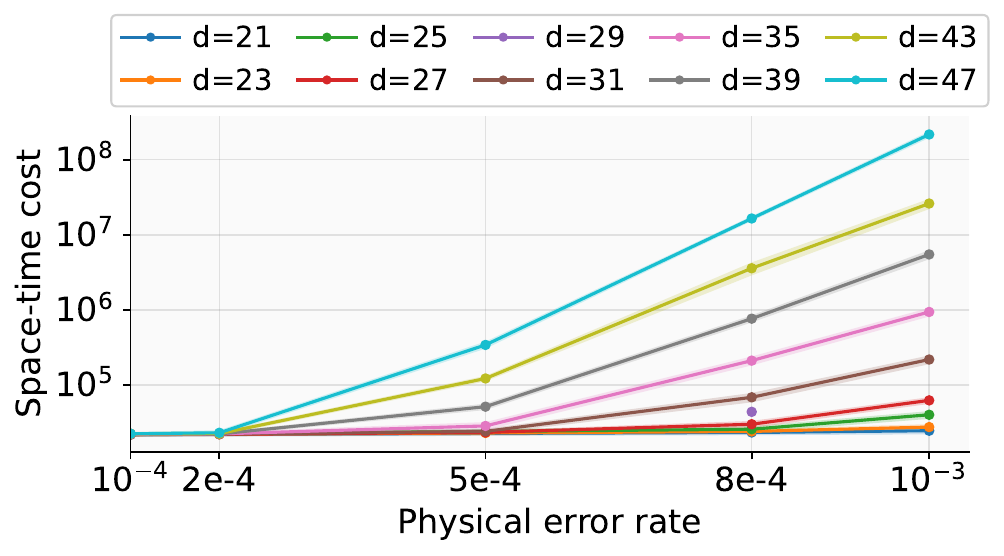}
        \caption{knn\_n25 for \Rz\ synthesis}
        \label{fig:rz_knn_sensitivity}
    \end{subfigure}
    \caption{Space–time cost as a function of PER, comparing different factory configurations for each production method}
    \vspace{-4.5pt}
    \label{fig:sensitivity}
\end{figure*}

However, stochastic analysis still provides additional useful insights. While the plateaus may coincide, the cycle count at the plateau is still higher under stochastic execution, confirming that the price is always present. More importantly for provisioning decisions, the peak per-cycle demand under stochastic execution is still lower than the deterministic peak. Provisioning factories based on $\gamma_{\text{peak}}$ still leads to over-provisioning. The factories needed to handle the worst-case deterministic cycle are never simultaneously demanded. The qubit savings from removing individual cultivation or Rz-synthesis units are modest (given their small footprint), but across an entire system with dozens of units, the cumulative savings can be repurposed. Stochastic simulation thus remains essential for right-sizing factory allocations across all three mechanisms, even when the cycle-time plateau is unaffected.

\subsection{Sensitivity to Code Distance and Physical Error Rate}
The optimal provisioning point depends not only on the number of factories but also on the factory configuration. Higher code distances yield higher-fidelity output states but increase production time and/or per-factory qubit cost; lower physical error rates reduce failure rates but are harder to achieve in hardware. We examine how the space-time cost landscape shifts across different configurations (Figure~\ref{fig:sensitivity}).
Increasing code distance raises the per-factory qubit cost and production time, shifting the space-time curve upward. Each additional factory is more expensive, so the penalty for over-provisioning grows, making accurate $F^*$ estimation more important at higher distances. The circuit takes longer to execute at any given~$F$ because each production round is slower, but the output states are of higher fidelity. Additionally, lowering the physical error rate reduces factory failure rates. At very low~$p$, the gap between deterministic and stochastic cycle counts narrows as factory aborts and stage failures become rare.

For distillation (Figure~\ref{fig:distil_sensitivity}), the space-time cost is nearly flat across PER for a given factory count, reflecting the protocol's low abort rate; the dominant axis of variation is factory count itself. Cultivation (Figure~\ref{fig:cultiv_sensitivity}) exhibits a different pattern where space-time cost is stable at low PER but rises sharply at higher $p$, with higher $d_2$ configurations diverging most severely. This reflects the compounding effect of stage failures since at high PER, the expected number of restarts grows rapidly, and larger $d_2$ values amplify the per-attempt cost of each failure. \Rz{} synthesis (Figure~\ref{fig:rz_knn_sensitivity}) shows the steepest sensitivity, with space-time cost spanning over two orders of magnitude across the PER range at high code distances. This is consistent with the cascading fixup behavior described before since each failed injection can trigger new production unit allocations, and at high PER these cascades become frequent.

Third, and most important is that the mechanisms differ not just in absolute cost but in their sensitivity to operating conditions. Distillation is resistant to reasonable PER variation but expensive in absolute terms; cultivation is cheap at low PER but fragile above a threshold; Rz synthesis offers low absolute cost at low PER and low distances but degrades fastest as conditions worsen. This means the choice of mechanism and the provisioning decision are coupled. This is a point that static analysis, which treats them independently, cannot capture. Our stochastic framework is most valuable in the near-term regime where physical error rates are highest and qubit budgets are tightest, precisely the regime where FTQC systems will first operate.

\section{Discussion}
\subsection{Effect of Circuit Structure}
The price and payoff are not uniform across circuits; their magnitude depends on two structural properties that architects should evaluate before committing to a provisioning strategy.

\subsubsection{Critical-path T-gate density}
Circuits in which a large fraction of non-Clifford gates lie on the critical path are most sensitive to injection errors. Each fixup on the critical path directly extends the schedule, so circuits with long serial chains of T gates experience compounding delays. In the extreme case of a fully serial T-gate sequence, the expected fixup-induced overhead grows linearly with T count. Our benchmarks confirm this; for circuits with high critical-path T-gate density (e.g. qft\_n18, where T gates are distributed across a long critical path) exhibit consistently higher overhead ratios in Figure~\ref{fig:overhead_cycles} than circuits with equivalent T counts but wider DAGs that absorb fixups off the critical path.

\subsubsection{Demand Burstiness}
Circuits with highly non-uniform per-layer T-gate counts exhibit sharper demand spikes under deterministic execution, and therefore benefit most from the smoothing payoff. A circuit whose T-gate demand is approximately uniform across layers gains little from demand smoothing because there are no pronounced spikes to attenuate. This is visible in Figure~\ref{fig:static_profiles} where knn\_n25 has a bursty profile (average 3.69, peak 15) and exhibits a 13.3\% peak reduction under stochastic execution, while qft\_n18 has a flatter profile (average 1.66, peak 10) and shows a correspondingly smaller reduction.
These two properties could serve as lightweight predictors to determine whether a given circuit will benefit significantly from stochastic-aware allocation or not, a direction we leave to future work.

\subsection{Comparison of Preparation Mechanisms}
Table~\ref{tab:mechanism_comparison} summarizes the key characteristics of each preparation mechanism along the dimensions that matter for provisioning. It highlights a fundamental architectural tradeoff: distillation offers predictable production timing but at high spatial cost, meaning over-provisioning wastes the most qubits and demand smoothing yields the largest savings. Cultivation and \Rz{} synthesis are spatially cheap but temporally variable, meaning the price (cycle-time overhead) is proportionally larger while the payoff (qubit savings per removed unit) is smaller. No single mechanism dominates across all operating points; the best choice depends on the target circuit's structure, the available qubit budget, and the tolerable execution time. What our analysis adds is that this choice cannot be made correctly without accounting for stochastic effects. In particular, a mechanism that appears superior under deterministic assumptions (e.g. cultivation for its low footprint) may exhibit higher-than-expected cycle overhead at moderate PER, shifting the balance.

\begin{table}
    \centering
    \caption{Magic-state preparation mechanisms}
    \label{tab:mechanism_comparison}
    \begin{tabular}{
    >{\centering\arraybackslash}p{2cm}
    >{\centering\arraybackslash}p{1.7cm}
    >{\centering\arraybackslash}p{1.7cm}
    >{\centering\arraybackslash}p{1.7cm}}
        \toprule
        & \textbf{Distillation} & \textbf{Cultivation} & \textbf{\Rz{} Synthesis} \\
        \midrule
        Factory Failure Probability & Low & High & High \\
        \addlinespace
        Spatial Footprint & High & Low & Low \\
        \addlinespace
        Production round length & Deterministic (if no aborts) & Variable (due to failures) & Variable (due to failures) \\
        \addlinespace
        Plateau shift under errors & Significant (early plateau) & Insignificant & Insignificant \\
        \addlinespace
        Fixup consequence & Not invoked again & Not invoked again & Invoked again if $R_z(2\theta)$ unavailable \\
        \bottomrule
    \end{tabular}
\end{table}

\subsection{The Space-Time-Fidelity Balancing Act}
\begin{figure}[t]
    \centering
    \includegraphics[width=0.75\linewidth]{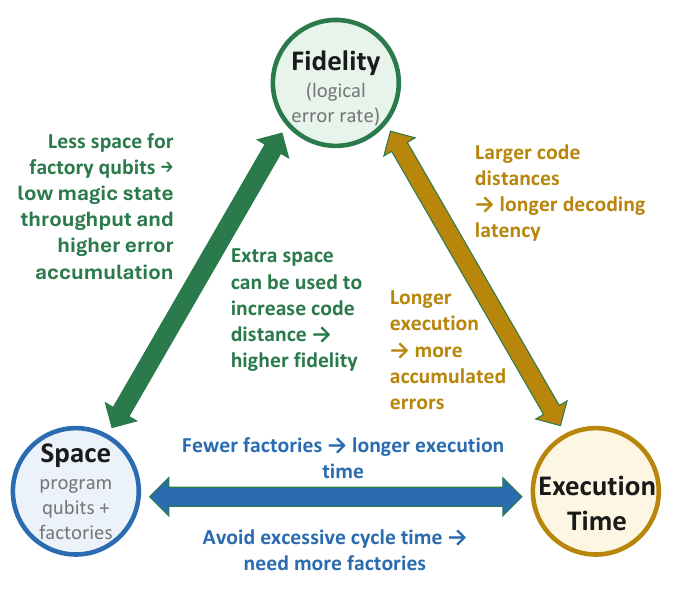}
    \caption{The Space-Time-Fidelity Balancing Act}
    \label{fig:tradeoff_discussion}
\end{figure}

Our analysis optimizes along two of the three axes that govern FTQC resource allocation: space (qubit count) and time (cycle count). The third axis, \emph{fidelity}, introduces an additional constraint. Longer execution times expose data qubits to idle errors, accumulating logical errors; conversely, reducing factory count frees qubits that could be reallocated to increase code distance, improving error correction. These two effects push in opposite directions, creating a three-way tension (Figure~\ref{fig:tradeoff_discussion}).

We do not model this fidelity dimension which may affect the choice of space-time-optimal $F^*$ once decoherence is included. However, our results provide a strictly tighter starting point than deterministic analysis. Since we show that deterministic provisioning over-allocates factories, any fidelity-aware optimization that begins from the deterministic estimate would inherit the over-allocation. Starting from our stochastic $F^*$ gives the fidelity optimizer a more accurate initial operating point and a larger unallocated qubit budget to work with. Incorporating idle-error tracking i.e. computing per-qubit idle durations during stall cycles and their effect on logical error rates, is a natural extension that would enable joint optimization across all three dimensions.

We do not model the routing and communication costs associated with teleporting a magic state from its production unit to the data qubit that consumes it. In surface-code architecture, this involves lattice-surgery operations whose latency depends on the spatial layout of factories relative to data qubits. Incorporating layout-dependent teleportation latency would introduce additional stochastic variability into the consumption side of the pipeline, likely amplifying both the price and the payoff effects we observe.

\section{Conclusion}
We have shown through stochastic simulation that non-determinism in fault-tolerant execution has a dual effect: it \emph{inflates} total execution time while \emph{deflating} peak per-cycle resource demand. Deterministic resource estimation therefore simultaneously underestimates execution time and overestimates factory requirements. For distillation, where each factory occupies hundreds to thousands of physical qubits, this over-provisioning is substantial; the stochastic space-time-optimal factory count $F^*$ is strictly lower than the deterministic optimum, and the freed qubits represent budget that can be redirected elsewhere. For cultivation and \Rz{} synthesis, where production units are smaller, the qubit savings per unit are modest, but the price of non-determinism is proportionally larger and the deterministic peak demand still overstates what is realized in practice.

Three takeaways emerge for FTQC system architects: 1) provisioning for deterministic peak demand is always wasteful, as the stochastic demand profile never reaches the deterministic peak, 2) the severity of both effects depends on circuit structure and operating regime, so provisioning decisions should be circuit-specific, and 3) most broadly, static resource estimation should be replaced by stochastic-aware simulation as the standard methodology for FTQC resource planning since the gap between deterministic predictions and realized execution is a systematic bias, not a second-order correction. Our framework provides the tool for this analysis; extending it to incorporate idle-error modeling, layout-aware teleportation latency, and scheduler–provisioning co-optimization would enable joint optimization across the full space-time-fidelity landscape, and we consider these natural next steps toward practical FTQC resource estimation.

\section*{Acknowledgements}
The authors acknowledge the Texas Advanced Computing Center (\href{http://www.tacc.utexas.edu}{TACC}) at The University of Texas at Austin for providing computational resources that have contributed to the research results reported in this paper. This work was funded in part by the Texas Quantum Institute (TQI), the NSF 24-599: Quantum Leap Challenge Institutes (QLCI), and the NSF STAQ project (PHY-2325080). Support is also acknowledged from the U.S. Department of Energy, Office of Science, National Quantum Information Science Research Centers, Quantum Systems Accelerator (Award No. DE-SCL0000121) and Office of Advanced Scientific Computing Research, Accelerated Research in Quantum Computing (Award No. DE-SC0025633). This research was, in part, funded by the U.S. Government. The views and conclusions contained in this document are those of the authors and should not be interpreted as representing the official policies, either expressed or implied, of the U.S. Government.

\bibliographystyle{IEEEtranS}
\bibliography{references}

@misc{Gidney_2024_cultivation,
      title={Magic state cultivation: growing T states as cheap as CNOT gates}, 
      author={Craig Gidney and Noah Shutty and Cody Jones},
      year={2024},
      eprint={2409.17595},
      archivePrefix={arXiv},
      primaryClass={quant-ph},
      url={https://arxiv.org/abs/2409.17595}, 
}

@article{Bravyi_2005,
   title={Universal quantum computation with ideal Clifford gates and noisy ancillas},
   volume={71},
   ISSN={1094-1622},
   url={http://dx.doi.org/10.1103/PhysRevA.71.022316},
   DOI={10.1103/physreva.71.022316},
   number={2},
   journal={Physical Review A},
   publisher={American Physical Society (APS)},
   author={Bravyi, Sergey and Kitaev, Alexei},
   year={2005},
   month=feb }

@inproceedings{Ding_2018,
   title={Magic-State Functional Units: Mapping and Scheduling Multi-Level Distillation Circuits for Fault-Tolerant Quantum Architectures},
   url={http://dx.doi.org/10.1109/MICRO.2018.00072},
   DOI={10.1109/micro.2018.00072},
   booktitle={2018 51st Annual IEEE/ACM International Symposium on Microarchitecture (MICRO)},
   publisher={IEEE},
   author={Ding, Yongshan and Holmes, Adam and Javadi-Abhari, Ali and Franklin, Diana and Martonosi, Margaret and Chong, Frederic},
   year={2018},
   month=oct, pages={828–840}}

@inproceedings{TamingBW_2017,
author = {Tannu, Swamit S. and Myers, Zachary A. and Nair, Prashant J. and Carmean, Douglas M. and Qureshi, Moinuddin K.},
title = {Taming the instruction bandwidth of quantum computers via hardware-managed error correction},
year = {2017},
isbn = {9781450349529},
publisher = {Association for Computing Machinery},
url = {https://doi.org/10.1145/3123939.3123940},
doi = {10.1145/3123939.3123940},
booktitle = {Proceedings of the 50th Annual IEEE/ACM International Symposium on Microarchitecture},
pages = {679–691},
numpages = {13},
location = {Cambridge, Massachusetts},
series = {MICRO-50 '17}
}

@inproceedings{Isailovic_2008,
author = {Isailovic, Nemanja and Whitney, Mark and Patel, Yatish and Kubiatowicz, John},
title = {Running a Quantum Circuit at the Speed of Data},
year = {2008},
isbn = {9780769531748},
publisher = {IEEE Computer Society},
address = {USA},
url = {https://doi.org/10.1109/ISCA.2008.5},
doi = {10.1109/ISCA.2008.5},
booktitle = {Proceedings of the 35th Annual International Symposium on Computer Architecture},
pages = {177–188},
numpages = {12},
series = {ISCA '08}
}

@article{gidney2021factor,
  title={How to factor 2048 bit RSA integers in 8 hours using 20 million noisy qubits},
  author={Gidney, Craig and Eker{\aa}, Martin},
  journal={Quantum},
  volume={5},
  pages={433},
  year={2021},
  publisher={Verein zur F{\"o}rderung des Open Access Publizierens in den Quantenwissenschaften}
}

@article{eastin2009restrictions,
  title={Restrictions on transversal encoded quantum gate sets},
  author={Eastin, Bryan and Knill, Emanuel},
  journal={Physical review letters},
  volume={102},
  number={11},
  pages={110502},
  year={2009},
  publisher={APS}
}

@article{akahoshi2024partially,
  title={Partially fault-tolerant quantum computing architecture with error-corrected clifford gates and space-time efficient analog rotations},
  author={Akahoshi, Yutaro and Maruyama, Kazunori and Oshima, Hirotaka and Sato, Shintaro and Fujii, Keisuke},
  journal={PRX quantum},
  volume={5},
  number={1},
  pages={010337},
  year={2024},
  publisher={APS}
}

@article{kubica2015universal,
  title={Universal transversal gates with color codes: A simplified approach},
  author={Kubica, Aleksander and Beverland, Michael E},
  journal={Physical Review A},
  volume={91},
  number={3},
  pages={032330},
  year={2015},
  publisher={APS}
}

@inproceedings{chadwick2024averting,
  title={Averting multi-qubit burst errors in surface code magic state factories},
  author={Chadwick, Jason D and Kang, Christopher and Viszlai, Joshua and Lin, Sophia Fuhui and Chong, Frederic T},
  booktitle={2024 IEEE International Conference on Quantum Computing and Engineering (QCE)},
  volume={1},
  pages={1089--1101},
  year={2024},
  organization={IEEE}
}

@article{holmes2019resource,
  title={Resource optimized quantum architectures for surface code implementations of magic-state distillation},
  author={Holmes, Adam and Ding, Yongshan and Javadi-Abhari, Ali and Franklin, Diana and Martonosi, Margaret and Chong, Frederic T},
  journal={Microprocessors and Microsystems},
  volume={67},
  pages={56--70},
  year={2019},
  publisher={Elsevier}
}

@inproceedings{qre1,
  title={Utilizing resource estimation for the development of quantum computing applications},
  author={Quetschlich, Nils and Soeken, Mathias and Murali, Prakash and Wille, Robert},
  booktitle={2024 IEEE International Conference on Quantum Computing and Engineering (QCE)},
  volume={1},
  pages={232--238},
  year={2024},
  organization={IEEE}
}

@article{qre2,
  title={Assessing requirements to scale to practical quantum advantage},
  author={Beverland, Michael E and Murali, Prakash and Troyer, Matthias and Svore, Krysta M and Hoefler, Torsten and Kliuchnikov, Vadym and Low, Guang Hao and Soeken, Mathias and Sundaram, Aarthi and Vaschillo, Alexander},
  journal={arXiv preprint arXiv:2211.07629},
  year={2022}
}

@article{bravyi2012magic,
  title={Magic-state distillation with low overhead},
  author={Bravyi, Sergey and Haah, Jeongwan},
  journal={Physical Review A—Atomic, Molecular, and Optical Physics},
  volume={86},
  number={5},
  pages={052329},
  year={2012},
  publisher={APS}
}

@article{terhal2015quantum,
  title={Quantum error correction for quantum memories},
  author={Terhal, Barbara M},
  journal={Reviews of Modern Physics},
  volume={87},
  number={2},
  pages={307--346},
  year={2015},
  publisher={APS}
}

@article{Litinski_2019_lattice_surgery,
   title={A Game of Surface Codes: Large-Scale Quantum Computing with Lattice Surgery},
   volume={3},
   ISSN={2521-327X},
   url={http://dx.doi.org/10.22331/q-2019-03-05-128},
   DOI={10.22331/q-2019-03-05-128},
   journal={Quantum},
   publisher={Verein zur Forderung des Open Access Publizierens in den Quantenwissenschaften},
   author={Litinski, Daniel},
   year={2019},
   month=mar, pages={128} }

@article{Litinski_2019_distillation,
   title={Magic State Distillation: Not as Costly as You Think},
   volume={3},
   ISSN={2521-327X},
   url={http://dx.doi.org/10.22331/q-2019-12-02-205},
   DOI={10.22331/q-2019-12-02-205},
   journal={Quantum},
   publisher={Verein zur Forderung des Open Access Publizierens in den Quantenwissenschaften},
   author={Litinski, Daniel},
   year={2019},
   month=dec, pages={205} }

@misc{li2022qasmbenchlowlevelqasmbenchmark,
      title={QASMBench: A Low-level QASM Benchmark Suite for NISQ Evaluation and Simulation}, 
      author={Ang Li and Samuel Stein and Sriram Krishnamoorthy and James Ang},
      year={2022},
      eprint={2005.13018},
      archivePrefix={arXiv},
      primaryClass={quant-ph},
      url={https://arxiv.org/abs/2005.13018}, 
}

@inproceedings{Sethi_2025, series={ASPLOS ’25},
   title={RESCQ: Realtime Scheduling for Continuous Angle Quantum Error Correction Architectures},
   url={http://dx.doi.org/10.1145/3676641.3716018},
   DOI={10.1145/3676641.3716018},
   booktitle={Proceedings of the 30th ACM International Conference on Architectural Support for Programming Languages and Operating Systems, Volume 2},
   publisher={ACM},
   author={Sethi, Sayam and Baker, Jonathan Mark},
   year={2025},
   month=mar, pages={1028–1043},
   collection={ASPLOS ’25} }

@misc{ross_2016_gridsynth,
      title={Optimal ancilla-free Clifford+T approximation of z-rotations}, 
      author={Neil J. Ross and Peter Selinger},
      year={2016},
      eprint={1403.2975},
      archivePrefix={arXiv},
      primaryClass={quant-ph},
      url={https://arxiv.org/abs/1403.2975}, 
}

@article{fowler_2012,
   title={Surface codes: Towards practical large-scale quantum computation},
   volume={86},
   ISSN={1094-1622},
   url={http://dx.doi.org/10.1103/PhysRevA.86.032324},
   DOI={10.1103/physreva.86.032324},
   number={3},
   journal={Physical Review A},
   publisher={American Physical Society (APS)},
   author={Fowler, Austin G. and Mariantoni, Matteo and Martinis, John M. and Cleland, Andrew N.},
   year={2012},
   month=sep }

@misc{matsumoto_2008,
      title={Representation of Quantum Circuits with Clifford and $\pi/8$ Gates}, 
      author={Ken Matsumoto and Kazuyuki Amano},
      year={2008},
      eprint={0806.3834},
      archivePrefix={arXiv},
      primaryClass={quant-ph},
      url={https://arxiv.org/abs/0806.3834}, 
}

@misc{giles_2019,
      title={Remarks on Matsumoto and Amano's normal form for single-qubit Clifford+T operators}, 
      author={Brett Giles and Peter Selinger},
      year={2019},
      eprint={1312.6584},
      archivePrefix={arXiv},
      primaryClass={quant-ph},
      url={https://arxiv.org/abs/1312.6584}, 
}

@misc{haner2017factoringusing2n2qubits,
      title={Factoring using 2n+2 qubits with Toffoli based modular multiplication}, 
      author={Thomas Häner and Martin Roetteler and Krysta M. Svore},
      year={2017},
      eprint={1611.07995},
      archivePrefix={arXiv},
      primaryClass={quant-ph},
      url={https://arxiv.org/abs/1611.07995}, 
}

@misc{yoshioka_transversal_2025,
    title = {Transversal gates for probabilistic implementation of multi-qubit {Pauli} rotations},
    url = {http://arxiv.org/abs/2510.08290},
    doi = {10.48550/arXiv.2510.08290},
    urldate = {2025-10-17},
    publisher = {arXiv},
    author = {Yoshioka, Nobuyuki and Seif, Alireza and Cross, Andrew and Javadi-Abhari, Ali},
    month = oct,
    year = {2025},
    note = {arXiv:2510.08290 [quant-ph]},
}

@misc{xu_distilling_2026,
    title = {Distilling {Magic} {States} in the {Bicycle} {Architecture}},
    url = {http://arxiv.org/abs/2602.20546},
    doi = {10.48550/arXiv.2602.20546},
    urldate = {2026-04-25},
    publisher = {arXiv},
    author = {Xu, Shifan and Liu, Kun and Rall, Patrick and He, Zhiyang and Ding, Yongshan},
    month = feb,
    year = {2026},
    note = {arXiv:2602.20546 [quant-ph]},
}

\end{document}